\documentclass[twocolumn,english,prl,showpacs,superscriptaddress,floatfix]{revtex4-1}
\usepackage[T1]{fontenc}
\usepackage[latin9]{inputenc}
\usepackage{babel}
\usepackage{amsmath}
\usepackage{amssymb}
\usepackage{graphicx}
\usepackage{esint}
\usepackage[colorlinks=true,citecolor=blue,linkcolor=blue,urlcolor=blue,breaklinks=true]{hyperref}

\makeatletter

\newcommand{\lyxdot}{.}

\@ifundefined{textcolor}{}
{%
 \definecolor{BLACK}{gray}{0}
 \definecolor{WHITE}{gray}{1}
 \definecolor{RED}{rgb}{1,0,0}
 \definecolor{GREEN}{rgb}{0,1,0}
 \definecolor{BLUE}{rgb}{0,0,1}
 \definecolor{CYAN}{cmyk}{1,0,0,0}
 \definecolor{MAGENTA}{cmyk}{0,1,0,0}
 \definecolor{YELLOW}{cmyk}{0,0,1,0}
 }

\usepackage{braket}

\makeatother

\begin{document}

\title{Constraints on measurement-based quantum computation in effective
cluster states}

\author{Daniel Klagges}

\email{klagges@fkt.physik.tu-dortmund.de}

\selectlanguage{english}%

\affiliation{Lehrstuhl für Theoretische Physik I, Otto-Hahn-Straße 4, TU Dortmund,
D-44221 Dortmund, Germany}

\author{Kai Phillip Schmidt}

\email{schmidt@fkt.physik.tu-dortmund.de}

\selectlanguage{english}%

\affiliation{Lehrstuhl für Theoretische Physik I, Otto-Hahn-Straße 4, TU Dortmund,
D-44221 Dortmund, Germany}
\begin{abstract}
The aim of this work is to study the physical properties of a one-way
quantum computer in an effective low-energy cluster state. We calculate
the optimal working conditions as a function of the temperature and of
the system parameters. The central result of our work is that any
effective cluster state implemented in a perturbative framework is
fragile against special kinds of external perturbations. Qualitative aspects of our work are 
important for any implementation of effective low-energy models containing
 strong multi-site interactions. 
\end{abstract}


\maketitle
A cluster state $\ket{\psi_{{\rm CS}}}$ is a quantum state defined
on some lattice (we focus on a square lattice) of qubits, which fulfills the following eigenvalue
equations 
\begin{equation}
K_{a}:=X_{a}\bigotimes_{b\in\Gamma(a)}Z{}_{b},\quad K_{a}\ket{\psi_{{\rm CS}}}=\ket{\psi_{{\rm CS}}}\quad,\label{eq:cluster stabiliser}
\end{equation}
 with $X_{a}$ and $Z_{b}$ being Pauli operators acting on qubits
$a$, $b$, and $\Gamma(a)$ is the set of nearest neighbors of site
$a$. So called one-way quantum computers (1-WQC) perform universal
quantum computations just by one-qubit measurements on a cluster state
\cite{raussendorf-2001}. While this saves the need to apply unitary
transformations to the quantum register, it requires to reliably prepare
a cluster state. One possibility is to cool the Hamiltonian 
\begin{equation}
H_{{\rm cl}}:=-\sum_{a\in C}K_{a}\label{eq:cluster hamilton}
\end{equation}
 into its non-degenerate ground state, which is by definition the
cluster state. A direct implementation of $H_{{\rm cl}}$ is not realistic,
as it contains multi-qubit interactions which are not realized in nature.

This suggests to search for a more realistic Hamiltonian with only
two-qubit interactions having the same non-degenerate ground state.
But such an Hamiltonian does not exist \cite{MichaelA2006147},
so one is limited to approximate the cluster state. One approach is
to use ancillary qubits to effectively mediate the many-qubit interactions. Unfortunately, this is not of practical use,
since the necessary precision scales with the system size \cite{nest2006graph}.

Alternatively, one can implement $H_{{\rm cl}}$ as an effective
low-energy model of a realistic Hamiltonian containing only two-qubit
interactions \cite{BartlettR,Griffin2008}. Two questions arise naturally:
What are the optimal working conditions to perform measurement-based
quantum computation (MBQC) in an effective cluster state? How robust are
effective cluster states with respect to external perturbations? In
the following, we show that any such effective cluster state, which
is implemented in a perturbative framework, is strongly affected by
external perturbations.

\emph{Model ---} We replace each qubit on a square lattice
by 4 physical qubits and we encode the logical cluster qubit
into the subspace defined by the projector $P~:=~\ket{0_{\log}}\braket{0000|+|1_{{\rm log}}}\bra{1111}$
\cite{BartlettR}. The
Hamiltonian $H:=gH_{0}+\lambda_{xz}V$ is then defined by: 
\begin{eqnarray*}
gH_{0} & := & -g\sum_{\mu\in\mathcal{L}}\sum_{i\leftrightarrow j}Z_{(\mu,i)}\otimes Z_{(\mu,j)},\\
\lambda_{xz}V & := & -\lambda_{xz}\sum_{\mu\in\mathcal{L}}\sum_{i=1}^{4}X_{(\mu,i)}\otimes Z_{\xi(\mu,i)}\quad.
\end{eqnarray*}

\begin{figure}
\begin{centering}
\includegraphics[width=0.44\columnwidth]{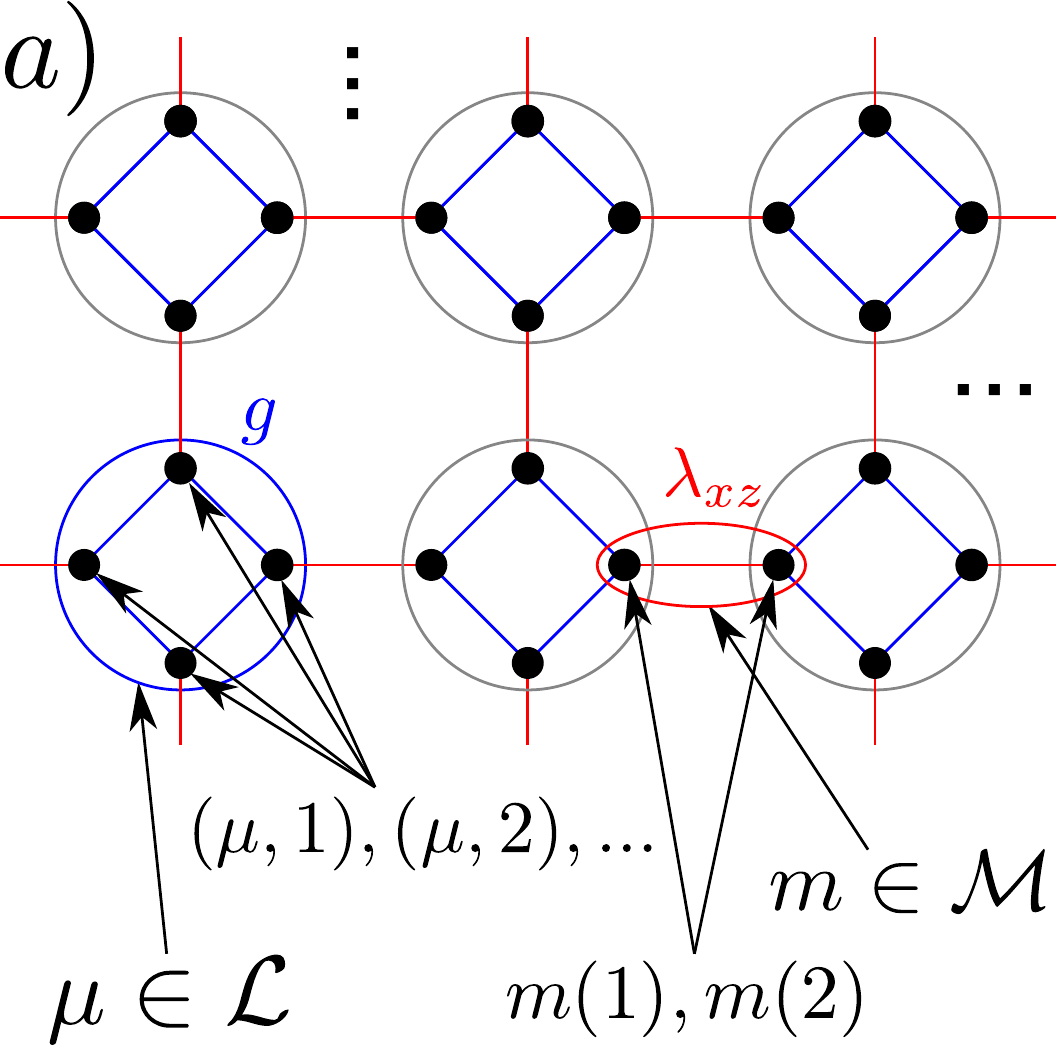}
\includegraphics[width=0.54\columnwidth]{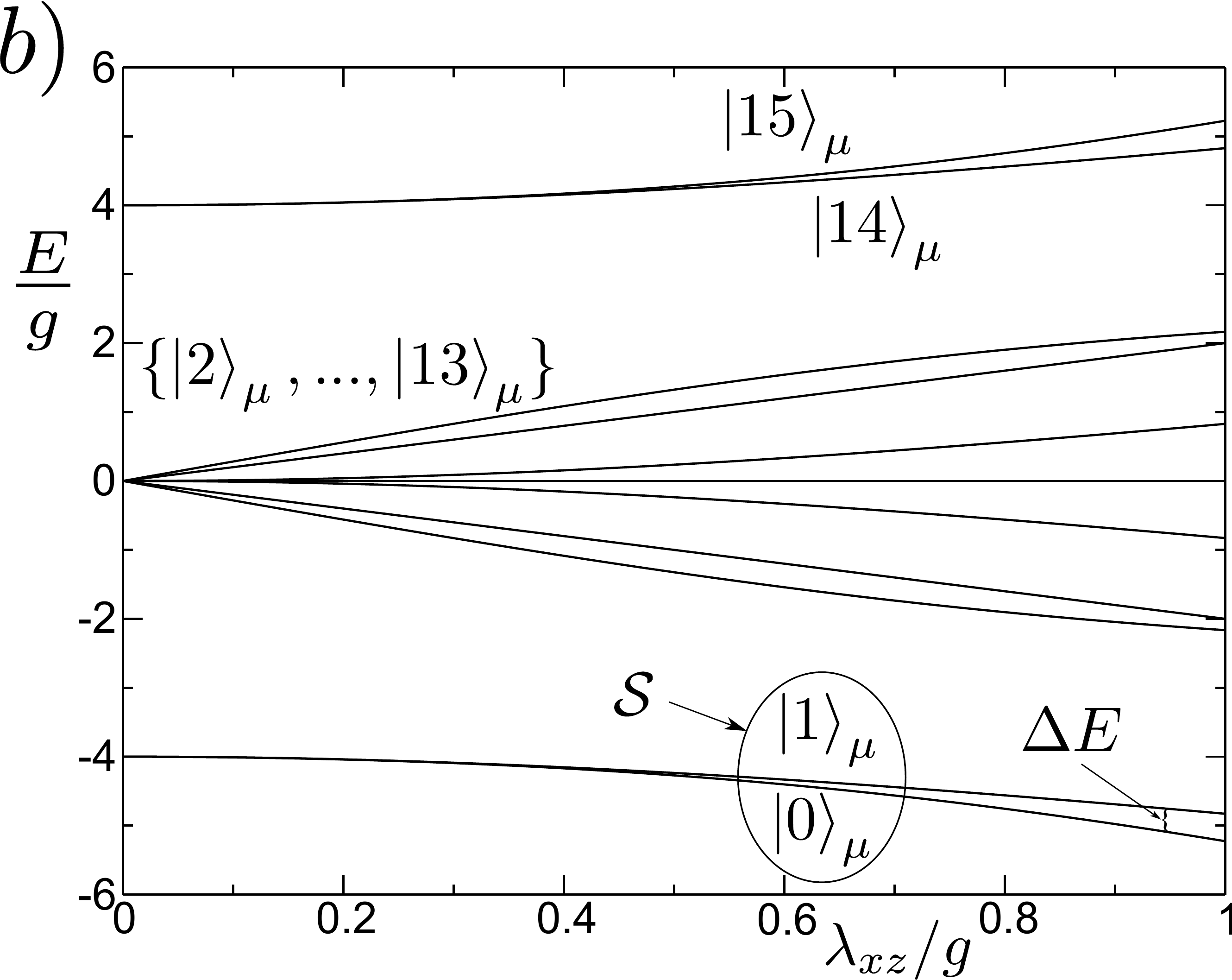}
\par\end{centering}

\caption{a) Physical qubits (black dots) on a CaVO lattice. The 4 physical
qubits of a lattice site $\mu\in\mathcal{L}$ (gray circles) are named
with the double indices $(\mu,1),...,(\mu,4)$. The two physical qubits
of a bond $m\in\mathcal{M}$ (red lines) are called $m(1)$ and $m(2)$.
To a physical qubit $(\mu,i)$, the neighboring qubit on the bond
is $\xi(\mu,i)$. The $ZZ$-interactions of $H_{0}$ are colored blue.
b) Energy spectrum of $H_{\mu}^{{\rm loc}}$ (Eq. \ref{eq:local ising}).\label{Flo:Physikalisches Quadratgitter} }
\end{figure}

The symbol $i\leftrightarrow j$ means, that the lattice sites $i$
and $j$ are related in some connected graph structure. Other notations
are illustrated in Fig.~\ref{Flo:Physikalisches Quadratgitter}a. The ground-state
space of $H_{0}$ is the space of all logical
qubits. 

It is possible to solve this model exactly 
by transforming $H$ into the base $\Sigma^{{\rm loc}}$ with the
base transformation $CZ_{\mathcal{M}}$ (controlled-$Z$ operators
on every bond) \cite{Griffin2008}. Using $(CZ_{\mathcal{M}})\, X_{(\mu,i)}\otimes Z_{\xi(\mu,i)}\,(CZ_{\mathcal{M}})=X_{(\mu,i)}\otimes I_{\xi(\mu,i)}$
and $[CZ_{\mathcal{M}},H_{0}]=0$, the transformed Hamiltonian reads
$H^{{\rm loc}}=\sum_{\mu\in\mathcal{L}}H{}_{\mu}^{{\rm loc}}$ with
\begin{eqnarray}
H_{\mu}^{{\rm loc}} & := & -g\sum_{i\leftrightarrow j}Z_{(\mu,i)}\otimes Z_{(\mu,j)}-\lambda_{xz}\sum_{i=1}^{4}X_{(\mu,i)}\quad.\label{eq:local ising}
\end{eqnarray}
 The $H_{\mu}^{{\rm loc}}$ are local transverse field Ising models
(TFIM) on each lattice site $\mu$ which can be solved by exact diagonalization.
If $R_{\mu}$ is the $16\times16$ matrix of the $16$ eigenvectors
$\ket{0}_{\mu}...\ket{15}_{\mu}$ of $H_{\mu}^{{\rm loc}}$, the diagonal
form of $H$ in the basis $\Sigma^{{\rm diag}}$ is: $H^{{\rm diag}}:=[\bigotimes_{\mu\in\mathcal{L}}R_{\mu}]H^{{\rm loc}}[\bigotimes_{\mu\in\mathcal{L}}R_{\mu}^{\dagger}]$
(see Fig.~\ref{Flo:Physikalisches Quadratgitter}b). Most importantly, the gap $\Delta E$ between
the (unique) ground state $\ket{\psi_{H}}$ and the first excited
state arises perturbatively in order 4 in $\lambda_{xz}/g$ \cite{Griffin2008,supMat}.

It is useful to generalize the cluster stabilizers $K_{a}$ into the
physical space 
\[
K_{\mu}:=\bigotimes_{i=1}^{4}X_{(\mu,i)}\otimes Z_{\xi(\mu,i)}\quad.
\]
 The $K_{\mu}$ can be also transformed into the basis $\Sigma^{{\rm loc}}$:
$K_{\mu}^{{\rm loc}}=(CZ_{\mathcal{M}})K_{\mu}(CZ_{\mathcal{M}})=\bigotimes_{i=1}^{4}X_{(\mu,i)}.$
In this basis it is easy to show, that the $K_{\mu}^{{\rm loc}}$
commute with each other and with $H^{\rm loc}$. Consequently, the eigenvalues $\pm 1$ for each stabilizer $K_{\mu}$
are conserved quantities. For $\lambda_{xz}\rightarrow\infty$, 
the ground state of $H^{{\rm loc}}$ is 
the polarized state which is eigenvector to all $K_{\mu}^{{\rm loc}}$ with eigenvalue
$+1$. The ground state of $H$ in the limit $\lambda_{xz}\rightarrow0$
is therefore the cluster state of the logical qubits.

The two low-energy states $\{\ket{0}_{\mu},\ket{1}_{\mu}\}$ of $H_{\mu}^{{\rm loc}}$ represent
an effective qubit on the according lattice site $\mu$ (see Fig.~\ref{Flo:Physikalisches Quadratgitter}b). We can therefore
derive an effective low-energy model
in the space $\mathcal{S}$ of all effective qubits 
\begin{equation}
\left.H^{{\rm diag}}\right|_{\mathcal{S}}=-\frac{\Delta E}{2}\sum_{\mu\in\mathcal{L}}\tilde{Z}_{\mu}=-\frac{\Delta E}{2}\sum_{\mu\in\mathcal{L}}\left.K_{\mu}^{{\rm diag}}\right|_{\mathcal{S}},\label{eq:effecitve n=00003D0000E4herung endliche xz}
\end{equation}
 where $\tilde{Z}_{\mu}$ being the Pauli $Z$-operator acting on
the effective qubit on the lattice site $\mu$ and $ K_{\mu}^{{\rm diag}} = R_{\mu}^{\dagger} K_{\mu}^{{\rm loc}} R_{\mu} $. It follows, that the effective low-energy approximation
of $H$ in the limit $\lambda_{xz}\rightarrow0$ is the cluster state
Hamiltonian $H_{{\rm cl}}$ of the logical qubits.

\emph{Fidelity} --- The usability of $H$ for quantum
computations depends on the question how {}``well'' the logical
cluster state is approximated by $\ket{\psi_{H}}$. This can be quantified
by the fidelity $F=|\braket{\psi_{H}|\psi_{\rm CS}}|^{2}$ of two states
and its generalization for density operators $F=\braket{\psi_{\rm CS}|\rho|\psi_{\rm CS}}$
\cite{springerlink:10.1007/s10701-009-9381-y}. The fidelity translates
to the {}``success probability{}`` of a MBQC using $\ket{\psi_{H}}$ as a resource state \cite{Griffin2008}.

As shown in \cite{Zhou2007}, for a Hamiltonian $H_{N}(\lambda)$
with control parameter $\lambda$, size $N$, and two ground states $\psi_{N}(\lambda),\psi_{N}(\lambda')$
it is: $\lim_{N\rightarrow\infty}F(\psi_{N}(\lambda),\psi_{N}(\lambda'))=d^{N}$
with $d\in[0,1]$ being constant. The {}``fidelity per site''
$d:=\lim_{N\rightarrow\infty}\sqrt[N]{F(\psi_{N}(\lambda),\psi_{N}(\lambda'))}$
is therefore intensive. Since $d<1$ for $\lambda_{xz}>0$,
the fidelity of the logical cluster state with $\ket{\psi_{H}}$ vanishes
for $N\rightarrow\infty$. This questions the usability for large
systems. However, the concept can still be applied by quantum error
correction techniques \cite{Griffin2008}. In this context the value
$d$ translates to the success probability per measurement which must
be large enough to fulfill the threshold theorem \cite{Knill2009}.

We calculate the fidelity $F(\ket{\psi_{{\rm CS}}},\rho)=\braket{\psi_{{\rm CS}}|\rho|\psi_{{\rm CS}}}$
of the logical cluster state $\ket{\psi_{{\rm CS}}}$ with the canonical
density operator $\rho:=\frac{1}{Z}e^{-\beta H}=\frac{1}{Z}\sum_{i}e^{-\beta E_{\ket{\psi_{i}}}}\ket{\psi_{i}}\bra{\psi_{i}}$. Here $Z~=~Tr(e^{-\beta H})$ denotes the partition function, $\beta=\frac{1}{k_{B}T}$, and one finds
\[
F(\ket{\psi_{{\rm CS}}},\rho)=F(\ket{+}_{\mathcal{L}},\rho_{\mathcal{L}}^{{\rm loc}})=F(\ket{+}_{\mu},\rho_{\mu}^{{\rm loc}})^{N}=:d^{N}.
\]
with $\ket{+}_{\mu}:=1/\sqrt{2}(\ket{0_{\rm log}}_{\mu}+\ket{1_{\rm log}}_{\mu}) $. Consequently, it is sufficient to study a single lattice site (we omit index $\mu$) 
\begin{eqnarray*}
d & = & \braket{+|R^{\dagger}\rho^{{\rm diag}}R|+}=\frac{1}{Z}\sum_{i=0}^{15}e^{-\beta E_{\ket{i}}}| \braket{i|R|+}|^{2},
\end{eqnarray*}
 with $Z=\sum_{i}e^{-\beta E_{\ket{i}}}$. Next, we approximate
$e^{-\beta E_{\ket{i}}}|\braket{i|R|+}|^{2}\approx0\quad\forall i\neq0$
and $Z~\approx~e^{-\beta E_{\ket{0}}}~+~e^{-\beta E_{\ket{1}}}$
which is justified by the following observations:

\textit{(a)} For MBQC, we have to choose the temperature
low enough so that even the first excited state plays a minor
role. Due to the exponential scaling factor we can omit all contributions
of high-energy states.

\textit{(b)} Due to the orthogonality of the vectors $\ket{i}$,
it is $|\braket{i|R|+}|^{2}\leq1-|\braket{0|R|+}|^{2}$ for
all $i\in\{1,..15\}$. For not too large $\lambda_{xz}$ we expect
$|\braket{0|R|+}|^{2}\lesssim1$, so the contributions of the
other states are small.

\textit{(c)} For the first excited state it is $|\braket{1|R|+}|^{2}=0$.
This is proven by the fact, that $R\ket{+}$ and $\ket{1}$ do
not have the same conserved eigenvalue of the $K$-operator.

\begin{figure}
\begin{centering}
\includegraphics[width=0.7\columnwidth]{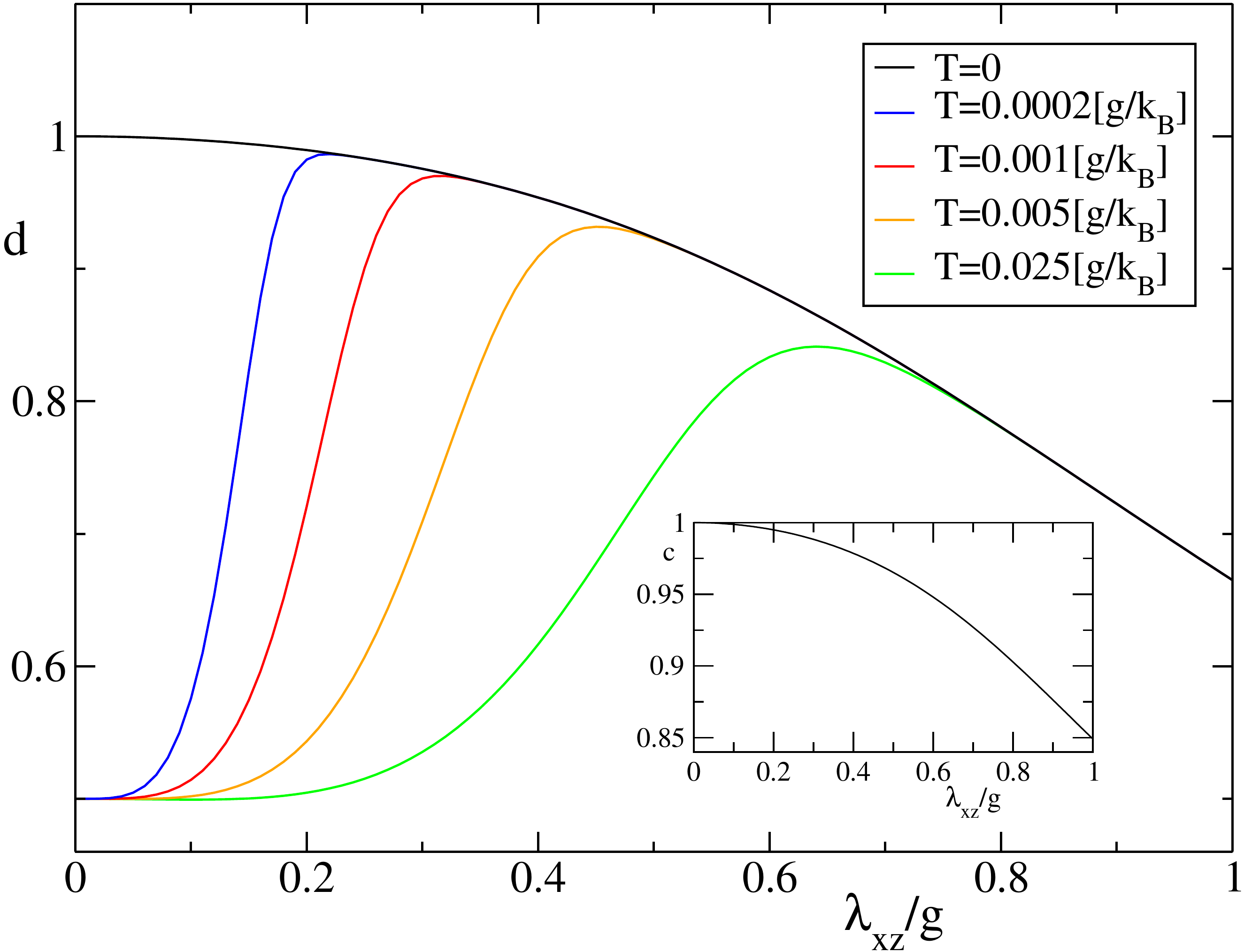} 
\par\end{centering}

\caption{Fidelity per site $d$ (Eq. \ref{eq:ungest=00003D0000F6rte fidelity pro gitterplatz})
in dependence of $\lambda_{xz}$ for different temperatures $T$. Inset: Contribution $c(\lambda_{xz}/g)$ of a physical $Z$-operator to the effective $\tilde{X}$-operator. \label{fig:Thermische-Fidelity-pro}}
\end{figure}

The resulting fidelity per site $d$ 
\begin{equation}
d\approx\frac{1}{1+e^{-\beta\Delta E}}|\braket{0|R|+}|^{2}.\label{eq:ungest=00003D0000F6rte fidelity pro gitterplatz}
\end{equation}
 is shown in Fig.~\ref{fig:Thermische-Fidelity-pro} for different
$T$. For finite $T$ and for small $\lambda_{xz}$, the fidelity
is dominated by thermal fluctuations. For large $\lambda_{xz}$ the
curve follows the zero-temperature ground-state fidelity. In between,
there is a trade off between both effects. If one assumes an error
correction algorithm for a 1-WQC with a simple error model of Pauli
errors %
\footnote{Strictly this does not cover the two Pauli operator errors of the
$V$-part, but we use it as a reference point. %
} as it is given in \cite{Raussendorf2006} with an error threshold
of $1.4\%$ ($d>0.986$), then the maximum $T_{{\rm max}}$ where
this threshold holds is $T_{{\rm max}}~=~2.18\,10^{-4}\, g/k_{{\rm B}}$.
It is reached for $\lambda_{xz}^{{\rm opt}}=0.222\, g$. 

The Hamiltonian $H$ can therefore be used as a 1-WQC under conditions
that in principle could be prepared in a laboratory. Next, we determine
the robustness of such an effective cluster state against external perturbations.

\emph{$Z$-field ---} First, we consider the presence of an external
field in $Z$-direction: 
\[
H_{z}:=gH_{0}+\lambda_{xz}V-h_{z}\sum_{\mu\in\mathcal{L}}\sum_{i=1}^{4}Z_{(\mu,i)}.
\]
 Let us formulate $H_{z}$ using the basis $\Sigma^{{\rm diag}}$ limited to $\mathcal{S} $. Since the perturbation commutes with $CZ_{\mathcal{M}}$, one has
\[
\left.H_{z}^{{\rm diag}}\right|_{\mathcal{S}}=-\frac{\Delta E}{2}\sum_{\mu\in\mathcal{L}}\tilde{Z}_{\mu}-4h_{z}c(\lambda_{xz}/g)\sum_{\mu\in\mathcal{L}}\tilde{X}_{\mu}\quad,
\]
where $c(\lambda_{xz}/g)\in\mathbb{R}$ can be read easily
from the matrix $R_{\mu}Z_{(\mu,i)}R_{\mu}^{\dagger}$.
One finds $\lim_{\lambda_{xz}\rightarrow0}c=1$ and the space $\mathcal{S} $ is decoupled from the high-energy space for this limit. 
The value $c(\lambda_{xz}/g)$ is plotted in Fig.~\ref{fig:Thermische-Fidelity-pro}.
 We stress that the scale $\Delta E$ is
of order 4 in $\lambda_{xz}$, while the scale of the $\tilde{X}$-field is
proportional to $h_{z}$. One therefore expects a polarization of
the ground state for very small ratios $h_{z}/\lambda_{xz}$.

\begin{figure}
\begin{centering}
\includegraphics[width=0.7\columnwidth]{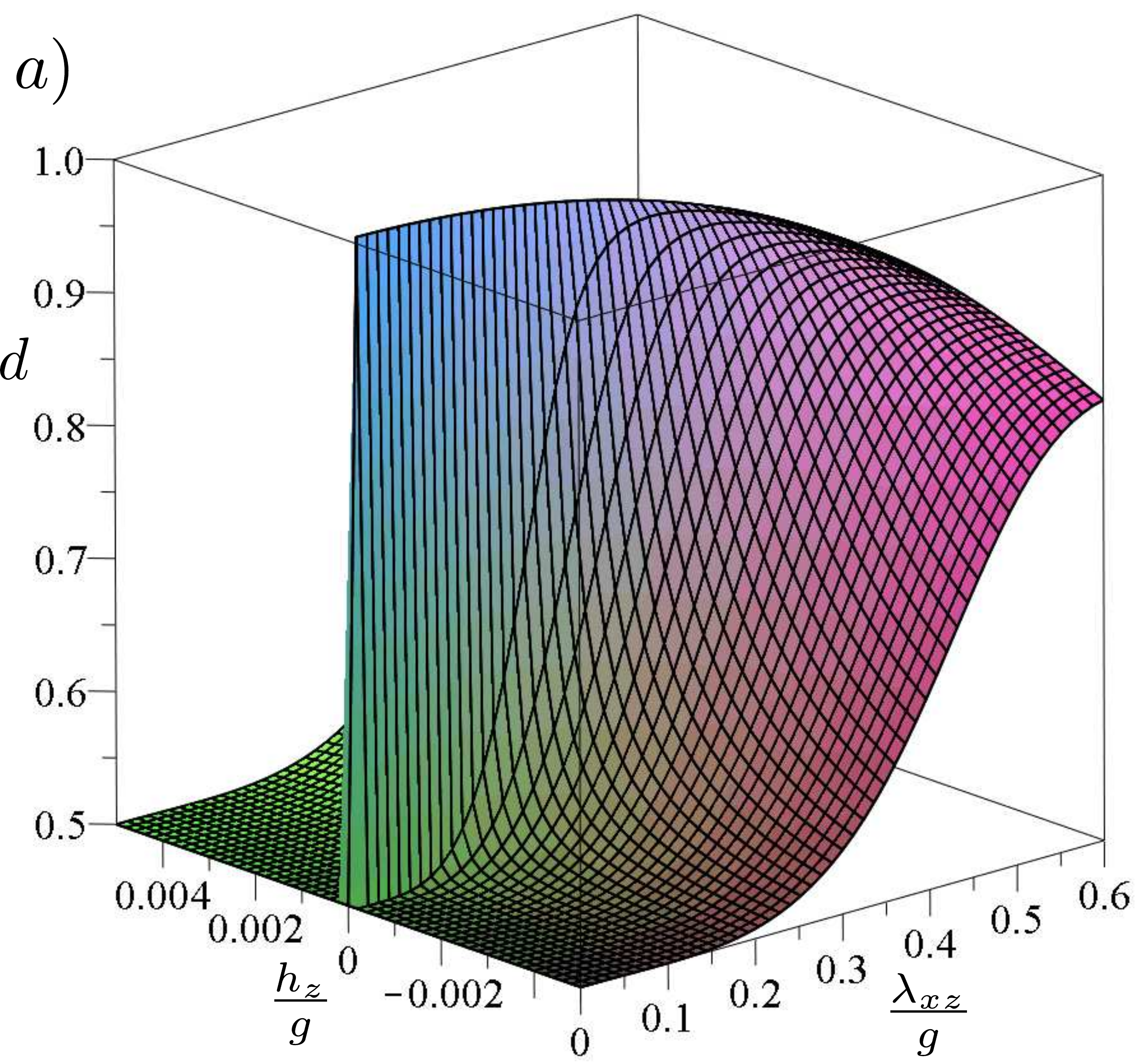} 
\par\end{centering}

\begin{centering}
\includegraphics[width=0.7\columnwidth]{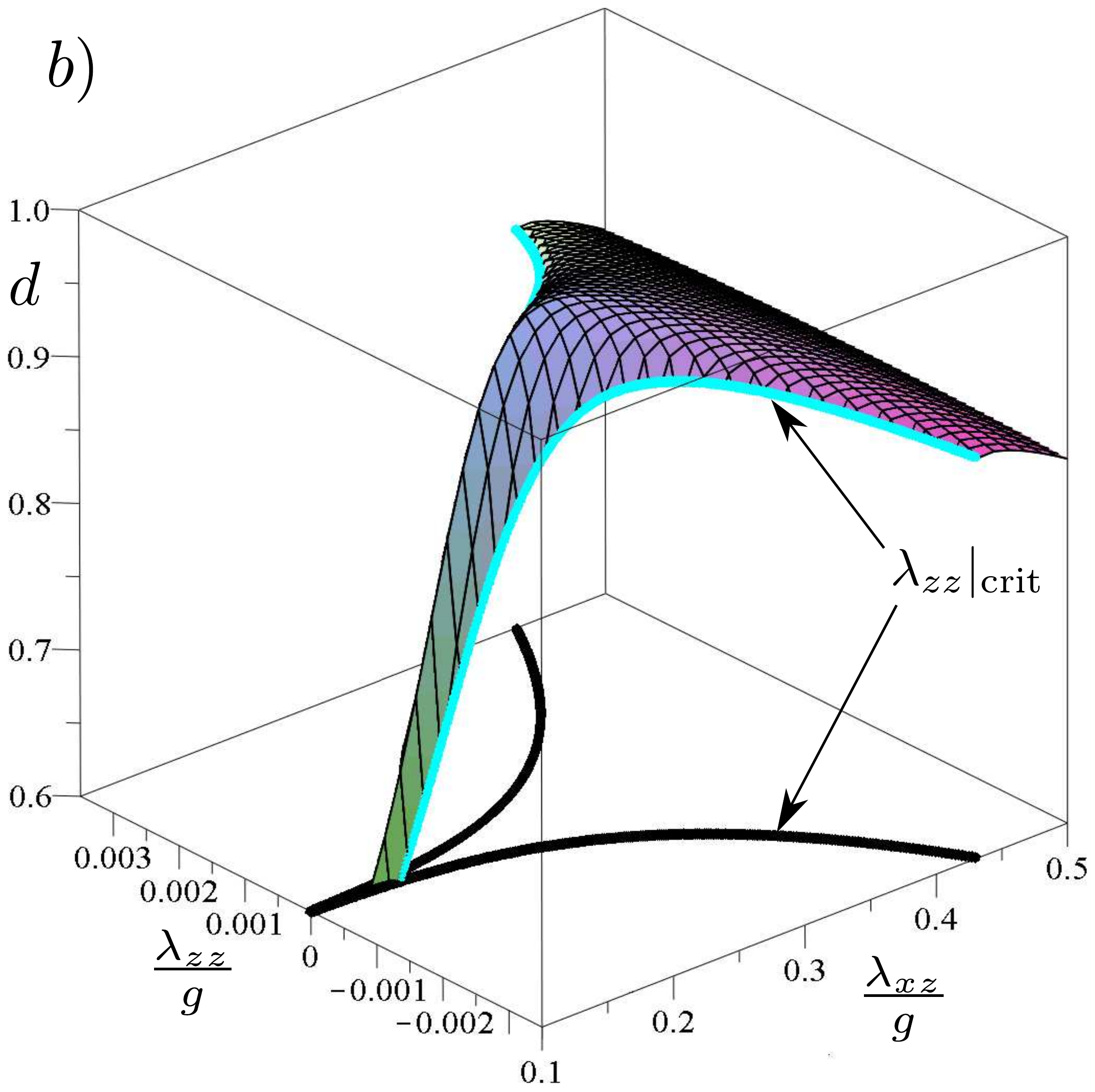} 
\par\end{centering}

\caption{\textit{a)} Ground-state fidelity per site $d=|\braket{0_{z}|R_{z}|+}|^{2}$ of $H_z$
as a function of $\lambda_{xz}$ and $h_{z}$ for $T=0$. \textit{b)}
Fidelity per site $d$ (Eq. \ref{eq:fidelity mit zz}) of $H_{zz}$ in
dependence of $\lambda_{xz}$ and $\lambda_{zz}$ for $T=0.001g/k_{{\rm B}}$.
\label{fig:Nicht-thermalisierte-Fidelity}}
\end{figure}

This is confirmed by solving the Hamiltonian $H_{z}$ exactly in the
basis $\Sigma^{{\rm loc}}$ 
\[
H{}_{z}^{{\rm loc}}=-\sum_{\mu\in\mathcal{L}}\left(H_{\mu}^{{\rm loc}}+h_{z}\sum_{i=1}^{4}Z_{(\mu,i)}\right)=:\sum_{\mu\in\mathcal{L}}H_{z,\mu}^{{\rm loc}}
\]
 which is still a sum of local terms $H_{z,\mu}^{{\rm loc}}$. The
fidelity per site of the ground state is calculated as in
the unperturbed case using the eigenvectors of $H_{z,\mu}^{{\rm loc}}$.
It can be seen in Fig.~\ref{fig:Nicht-thermalisierte-Fidelity}a that
already very small ratios $h_{z}/\lambda_{xz}$ have a significant
impact on $d$. For the above example, one finds the upper
bound $h_{z}^{{\rm max}}=1.52\,10^{-5}\, g$ satisfying the threshold
$d\geq0.986$ at $T=0$. Thermal fluctuations play only a minor role, because the gap is strongly
increased by the external field \cite{supMat}.

\emph{$ZZ$-coupling ---} Second, we consider the effect of additional
Ising $ZZ$-couplings on the bonds $m\in\mathcal{M}$ 
\[
H_{zz}:=gH_{0}+\lambda_{xz}V-\lambda_{zz}\sum_{m\in\mathcal{M}}Z_{m(1)}Z_{m(2)}.
\]
 Now we formulate the Hamiltonian using the basis $\Sigma^{{\rm diag}}$ limited to $\mathcal{S}$
\[
\left.H_{zz}^{\rm diag}\right|_{\mathcal{S}} =-\frac{\Delta E}{2}\sum_{\mu\in\mathcal{L}}\tilde{Z}_{\mu}-\lambda_{zz}c^2(\lambda_{xz})\sum_{m\in M}\tilde{X}_{m(1)}\tilde{X}_{m(2)},
\]
where again the space $\mathcal{S} $ is decoupled from the high-energy space for the limit $\lambda_{xz}\rightarrow0$.
 This Hamiltonian represents a
TFIM on the square lattice. For this model a quantum phase transition
takes place at $\frac{2\lambda_{zz}|_{{\rm crit}} c^2}{\Delta E}~=~0.3285$
separating an ordered from a disordered phase \cite{0305-4470-23-10-018}. The gap closes at the critical point
changing the ground state significantly,
so the system is not useful for MBQC anymore.
 The energy $\Delta E$ is a 4th-order term in $\lambda_{xz}$, while the Ising part 
is of the order $\lambda_{zz}$. Very small values $\lambda_{zz}/\lambda_{xz}$ are therefore sufficient
to destroy the cluster phase. The estimated critical line $\lambda_{zz}|_{{\rm crit}}$
is shown in Fig.~\ref{fig:Nicht-thermalisierte-Fidelity}b
as a function of $\lambda_{xz}$.

To approximately calculate the fidelity per site for finite
temperatures we use the analog of Eq.~\ref{eq:ungest=00003D0000F6rte fidelity pro gitterplatz}:
\begin{equation}
d\approx\frac{1}{1+\frac{1}{4\pi^{2}}\iintop_{\vec{k}}e^{-\beta\omega(\vec{k})}d\vec{k}}d(|\psi_{H_{zz}|_{\mathcal{S}}}\rangle,|\psi_{{\rm CS}}|_{\mathcal{S}}\rangle),\label{eq:fidelity mit zz ohne xz}
\end{equation}
 We note that point \textit{(c)} is no longer valid as $K$
is no longer conserved, but the use of Eq.~\ref{eq:ungest=00003D0000F6rte fidelity pro gitterplatz}
is still justified by \textit{(a)} and \textit{(b)}. The dispersion $\omega(\vec{k})\approx\sqrt{(\Delta E_{zz})^{2}+(v\cdot|\vec{k}|)^{2}} $ of the first excited mode is taken into account by this equation, while correlated excitations are neglected \cite{PhysRevB.50.13515}.
The energy gap $\Delta E_{zz}$ of the TFIM is calculated by a dlogPad\'e~{[}6,6{]}
approximation of its order 13 series expansion \cite{0305-4470-23-10-018} and $v=0.99\,\Delta E/2$ \cite{0305-4470-33-38-303} is the spin wave velocity at the critical point.

We now transform $\ket{\psi_{{\rm CS}}}$ into the effective basis
\begin{equation}
\left.|\psi_{{\rm CS}}^{{\rm diag}}\rangle\right|_{\mathcal{S}} = \bigotimes_{\mu\in\mathcal{L}}\left.R_{\mu}|+_{\mu}\rangle\right|_{\mathcal{S}}
 = \bigotimes_{\mu\in\mathcal{L}}|0_{\mu}\rangle\braket{0_{\mu}|R_{\mu}|+_{\mu}}
\end{equation}
 (using $ \braket{1|R_{\mu}|+_{\mu}}=0 $) such that Eq.~\ref{eq:fidelity mit zz ohne xz} reads 
\begin{equation}
d \approx \frac{|\braket{0|R|+}|^{2}}{1+\frac{1}{2\pi}\intop_{0}^{2\sqrt{\pi}}e^{-\beta\omega(r)}r\, dr}\,\,\ d_{{\rm TFIM}}\label{eq:fidelity mit zz}
\end{equation}
 where $d_{{\rm TFIM}}:=d(\ket{\psi_{H_{zz}|_{\mathcal{S}}}},\bigotimes_{\mu\in\mathcal{L}}\ket{0_{\mu}})$
corresponds to the ground-state fidelity per site of a TFIM with the
polarized state. We have calculated it as a high-order series expansion about the high-field limit 
\begin{eqnarray*}
d_{{\rm TFIM}} & = & 1-\frac{1}{8}\lambda^{2}-{\frac{93}{256}}\lambda^{4}-{\frac{2961}{2048}}{\it \lambda}^{6}-{\frac{243005}{32768}}{\it \lambda}^{8}\\
 &  & -{\frac{812949139}{18874368}}{\it \lambda}^{10}-{\frac{17716040461601}{65229815808}}{\it \lambda}^{12}
\end{eqnarray*}
where $\lambda:=\frac{2\lambda_{zz} c^2}{\Delta E}$ \cite{supMat}.

The fidelity per site of $H_{zz}|_{\rm eff}$ is plotted in Fig.~\ref{fig:Nicht-thermalisierte-Fidelity}b.
One sees, that close to the critical
point the fidelity drops due to quantum and thermal fluctuations.
For the above example, one finds the upper
bound $\lambda_{zz}^{{\rm max}}=8.33\,10^{-5}\, g$ satisfying the threshold
$d\geq0.986$ at $T=0$.

We additionally calculated the energy gap and the ground state fidelity of the full Hamiltonian $H_{zz}$ as series expansions.
The high-energy contributions turned out to be negligible corrections to the low-energy results \cite{supMat}.

\emph{Conclusions} --- We have seen that the effective cluster state
of $H$ could be used as a 1-WQC under conditions that in principle
can be prepared in a laboratory. But to be of practical use, effective
cluster states must be also robust against additional perturbations.

We have shown that already very small external perturbations can have
a significant impact on effective cluster states. Typically, the
effective multi-site interactions yielding the effective cluster state
arise in a high order in perturbation theory (here order 4).
As a consequence, any external perturbation acting in the effective
low-energy model in a lower order (here order 1) represents
a strong constraint for the effective implementation of a 1-WQC. 
This effect is present on any lattice and in any dimension for the problem
 studied in this work.

The physical mechanism leading to the dramatic loss of fidelity is
actually very different for the two perturbations we have considered.
The external $Z$-field leads to a polarization of the ground state
and therefore to a reduction of entanglement. The additional Ising
coupling induce thermal and quantum fluctuations due to a quantum phase transition. 

The qualitative aspects of our work are relevant for a much broader 
class of problems: any effective low-energy model which is derived 
perturbatively and which contains dominant multi-site interactions
 is expected to be affected by external perturbations. The physical 
reason is that effective $n$-site interactions arise typically in 
order $n$ while it is likely that external perturbations exist which
 act non-trivially on the effective low-energy degrees of freedom 
already in a lower order.
 
A prominent example is Kitaev's honeycomb model which contains the so-called 
toric code as an effective low-energy model perturbatively in order 4 \cite{Kitaev20062}. The toric code is a topological stabilizer code consisting solely of 4-spin 
interactions. One can easily show that exactly the same kind of external perturbations studied in this work give again rise to operators in the effective model in order one perturbation theory causing a breakdown of the topological phase for small external perturbations. 
   
In the light of the severe constraints found in this work for the realization of effective cluster states, let us finally mention concepts for MBQC using elementary entities with larger spins which represents a promising route for future research \cite{Chen2008,PhysRevLett.101.010502,PhysRevA.82.052309,PhysRevA.84.042333}.
\begin{acknowledgments}
K.P.S. acknowledges ESF and EuroHorcs for funding through his EURYI.\end{acknowledgments}

\newpage{}
\onecolumngrid

\section*{Supplemental Material}

\subsection*{Spectral properties of $H$}

The spectrum of $H$ (and $H^{{\rm diag}}$) is the product of the spectra of the local $H_{\mu}^{{\rm loc}}$
terms. In Tab.~\ref{tab:Eigenenergieen-der-Eigenzust=0000E4nde}
one can see the eigenenergies of the $H_{\mu}^{{\rm loc}}$ eigenvectors
$\ket{0}_{\mu},...,\ket{15}_{\mu}$.
\begin{table}[b]
\begin{centering}
\begin{tabular}{|c|c|}
\hline 
State & $E_{\ket{i}_{\mu}}$\tabularnewline
\hline 
\hline 
$\ket{0}_{\mu}$ & $-2g\,\sqrt{2+2\,{\frac{\lambda_{xz}^{2}}{g^{2}}}+2\,\sqrt{{\frac{\lambda_{xz}^{4}}{g^{4}}}+1}}$\tabularnewline
\hline 
$\ket{1}_{\mu}$ & $-2g-2g\,\sqrt{1+{\frac{\lambda_{xz}^{2}}{g^{2}}}}$\tabularnewline
\hline 
$\ket{2}_{\mu}$ & $-2g\,\sqrt{2+2\,{\frac{\lambda_{xz}^{2}}{g^{2}}}-2\,\sqrt{{\frac{\lambda_{xz}^{4}}{g^{4}}}+1}}$\tabularnewline
\hline 
$\ket{3}_{\mu},\ket{4}_{\mu}$ & $-2g\,\lambda_{xz}$\tabularnewline
\hline 
$\ket{5}_{\mu}$ & $2g-2g\,\sqrt{1+{\frac{\lambda_{xz}^{2}}{g^{2}}}}$\tabularnewline
\hline 
$\ket{6}_{\mu},\ket{7}_{\mu},\ket{8}_{\mu},\ket{9}_{\mu}$ & $0$\tabularnewline
\hline 
$\ket{10}_{\mu}$ & $-2g+2g\,\sqrt{1+{\frac{\lambda_{xz}^{2}}{g^{2}}}}$\tabularnewline
\hline 
$\ket{11}_{\mu},\ket{12}_{\mu}$ & $2g\,\lambda_{xz}$\tabularnewline
\hline 
$\ket{13}_{\mu}$ & $2g\,\sqrt{2+2\,{\frac{\lambda_{xz}^{2}}{g^{2}}}-2\,\sqrt{{\frac{\lambda_{xz}^{4}}{g^{4}}}+1}}$\tabularnewline
\hline 
$\ket{14}_{\mu}$ & $2g+2g\,\sqrt{1+{\frac{\lambda_{xz}^{2}}{g^{2}}}}$\tabularnewline
\hline 
$\ket{15}_{\mu}$ & $2g\,\sqrt{2+2\,{\frac{\lambda_{xz}^{2}}{g^{2}}}+2\,\sqrt{{\frac{\lambda_{xz}^{4}}{g^{4}}}+1}}$\tabularnewline
\hline 
\end{tabular}
\par\end{centering}

\caption{Eigenenergies of the $H_{\mu}^{{\rm loc}}$ eigenvectors $\ket{0}_{\mu},...,\ket{15}_{\mu}$\label{tab:Eigenenergieen-der-Eigenzust=0000E4nde}}
\end{table}
 The unique ground state 
\[
\ket{\psi_{H}}=\bigotimes_{\mu\in\mathcal{L}}\ket{0}_{\mu}
\]
 of $H^{{\rm diag}}$ is characterized by all lattice sites being
in the $\ket{0}_{\mu}$ state. The first excited space is $N$-fold
degenerated and is spanned by the states ($\mu\in\mathcal{L}$)
\[
\ket{\psi_{{\rm EX}}^{(\mu)}}=\ket{1}_{\mu}\bigotimes_{\nu\in\mathcal{L}/\{\mu\}}\ket{0}_{\nu}.
\]
For the energy gap $\Delta E=E_{\ket{1}_{\mu}}-E_{\ket{0}_{\mu}}$
between the ground state and the ($N$-fold degenerated) first excited
space one therefore finds 
\[
\Delta E=-2g\left(1+\sqrt{1+{\frac{\lambda_{xz}^{2}}{g^{2}}}}-\sqrt{2+2{\frac{\lambda_{xz}^{2}}{g^{2}}}+2\sqrt{{\frac{\lambda_{xz}^{4}}{g^{4}}}+1}}\right).
\]
It is independent of the system size. From the Taylor expansion of
$\Delta E$ one sees, that it is of order 4 in $\lambda_{xz}$
\[
\Delta E={\frac{5}{8}}{\left(\frac{\lambda_{xz}}{g}\right)}^{4}-{\frac{7}{32}}{\left(\frac{\lambda_{xz}}{g}\right)}^{6}-{\frac{21}{512}}{\left(\frac{\lambda_{xz}}{g}\right)}^{8}-{\frac{33}{2048}}{\left(\frac{\lambda_{xz}}{g}\right)}^{10}+O\left({\left(\frac{\lambda_{xz}}{g}\right)}^{12}\right).
\]

\begin{figure}
\begin{centering}
\includegraphics[width=0.6\columnwidth]{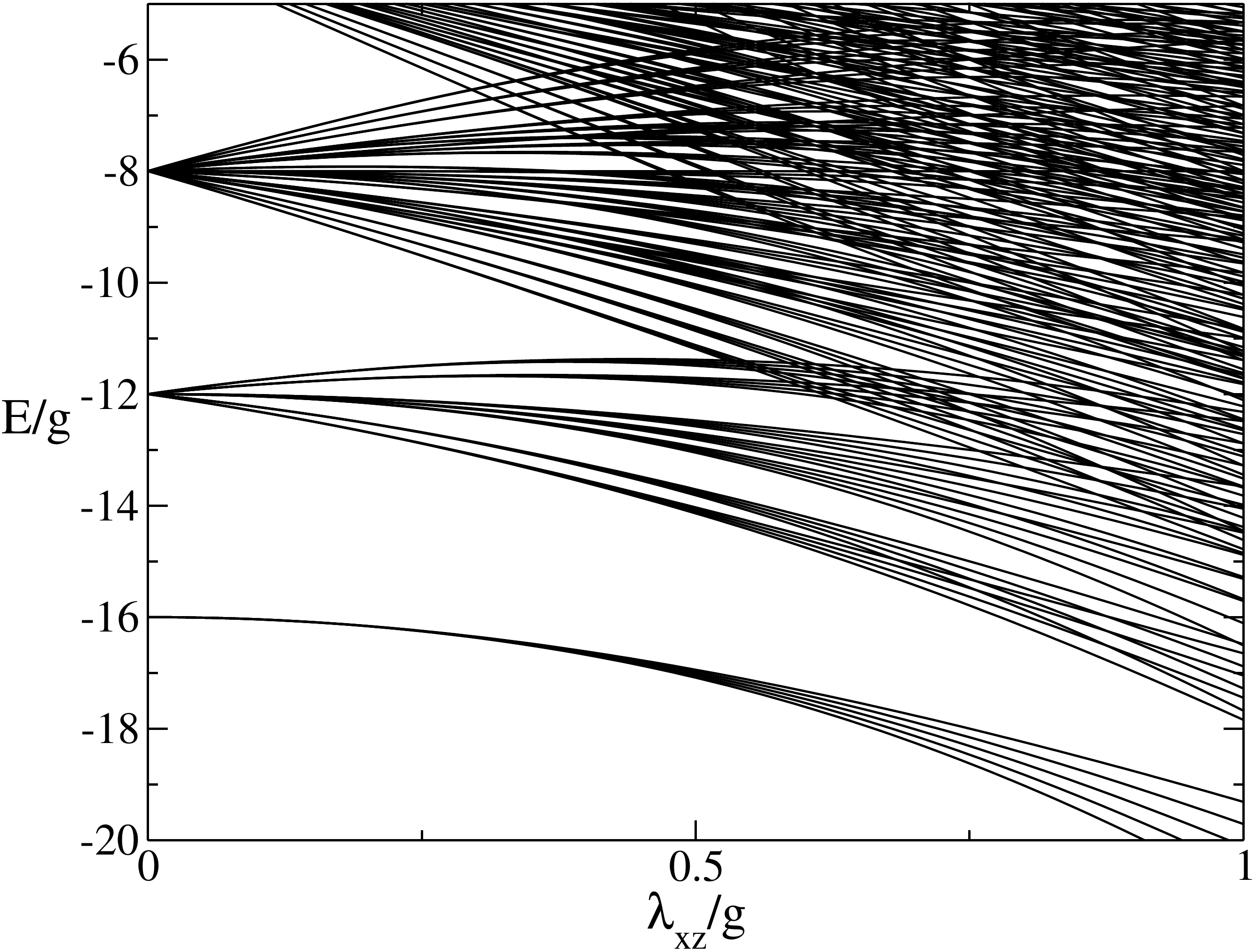}
\par\end{centering}

\caption{Lower part of the spectrum of $H$ for $N=4$\label{fig:Low-spectrum-of}}

\end{figure}

All of the first $N$ excited spaces can be characterized by some
lattice sites being in the $\ket{1}_{\mu}$ state and all other lattice
sites being in the $\ket{0}_{\mu}$ state. They form the space $\mathcal{S}$
of the effective qubits. The $n$-th excited space is $\binom{N}{n}$
times degenerated and the gap between the $n$-th and the $(n+1)$-th
space is $\Delta E$. All high-energy states are characterized by
at least one lattice site being in one of the states $\ket{2},...,\ket{15}$.
In Fig.~\ref{fig:Low-spectrum-of} you can see the lower part of
the spectrum for the example $N=4$.

\subsection*{Thermal fidelity of $H_{z}$}

To calculate the fidelity per site for finite temperatures for the
case of an external $Z$-field one must note, that the points \textit{(b)}
and \textit{(c)}\textit{\emph{ are no longer valid to justify the
approximation, }}since the ground-state fidelity quickly decreases
with $h_{z}$ and since $K$ is no longer a conserved quantity. Because
of \emph{(a)} it still is possible to approximate $e^{-\beta E_{\ket{i_{z}}}}|\braket{i_{z}|R_{z}|+}|^{2}\approx0\quad\forall i\geq2$
and $Z(\beta)\approx e^{-\beta E_{\ket{0_{z}}}}+e^{-\beta E_{\ket{1_{z}}}}$,
where $R_{z}$ is the $16\times16$ Matrix of the $H_{z,\mu}^{{\rm loc}}$
eigenvectors $\ket{0_{z}},...,\ket{15_{z}}$. One finds
\begin{equation}
d\approx\frac{1}{1+e^{-\beta\Delta E_{z}}}|\braket{0_{z}|R_{z}|+}|^{2}+\frac{1}{1+e^{\beta\Delta E_{z}}}|\braket{1_{z}|R_{z}|+}|^{2}.\label{eq:Fidelity with hz}
\end{equation}

\begin{figure}[p]
\begin{centering}
\includegraphics[width=0.48\linewidth]{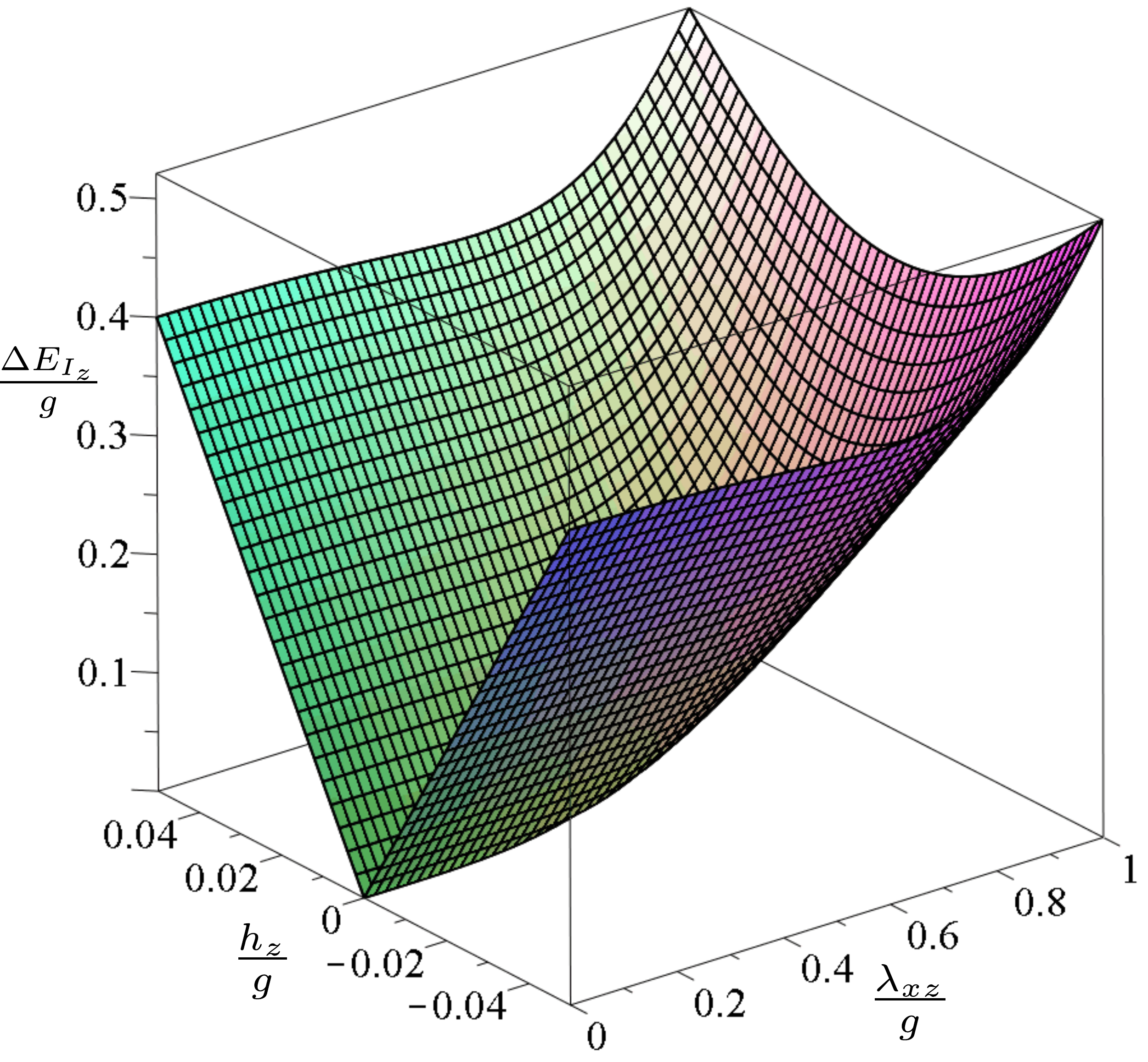}\hspace{0.03\linewidth}\includegraphics[width=0.48\linewidth]{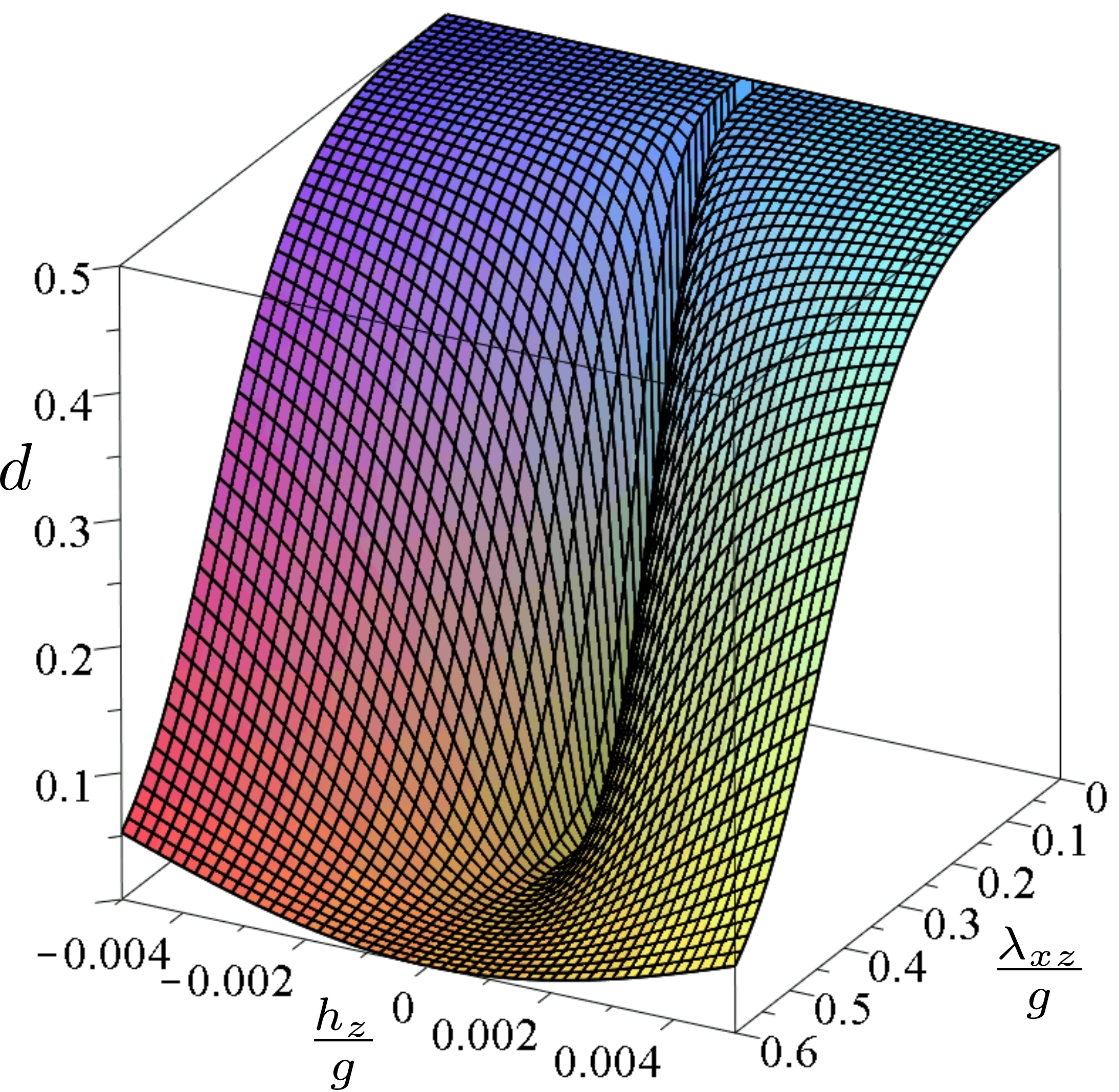}\bigskip{}
\includegraphics[width=0.48\linewidth]{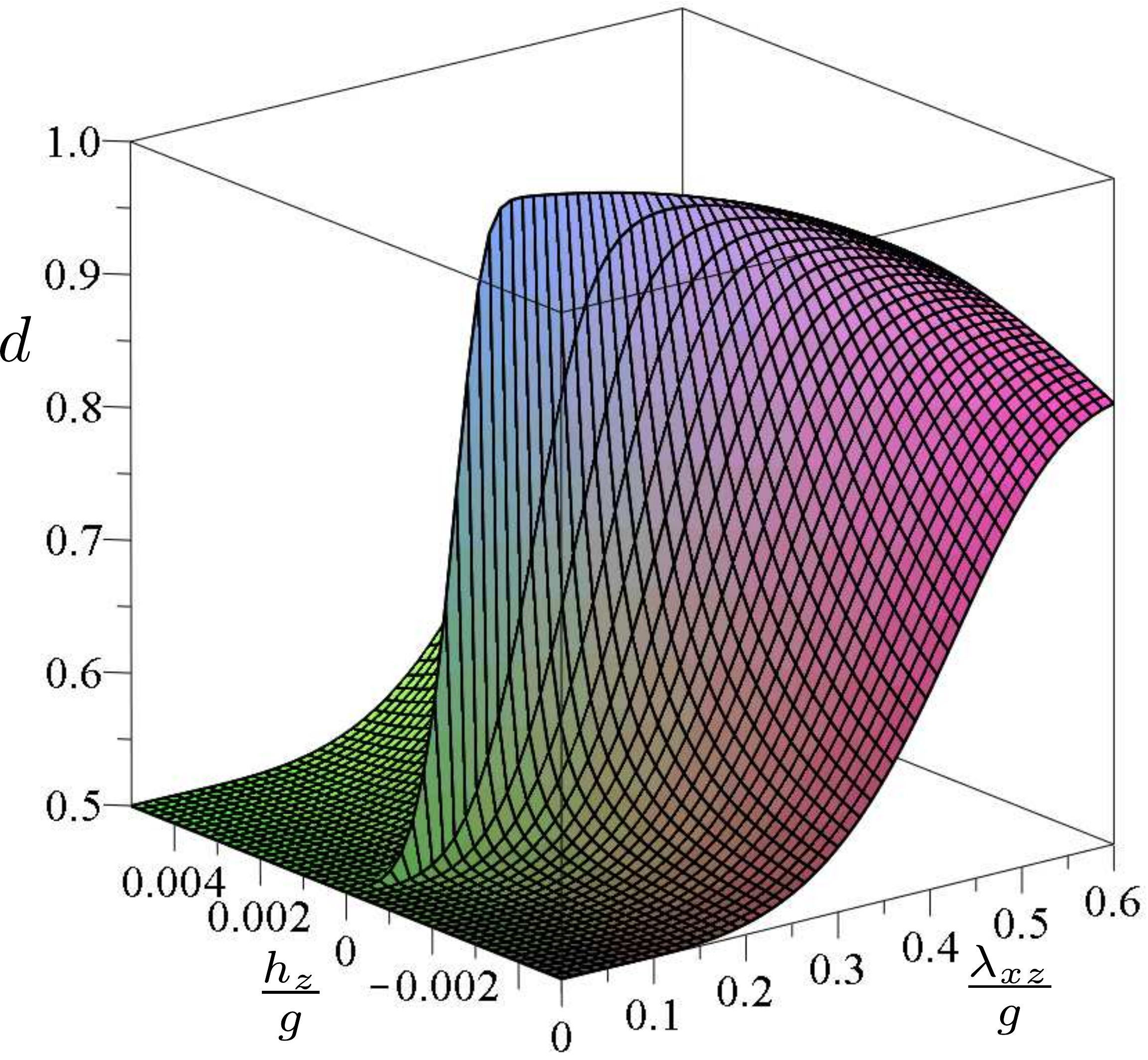}\hspace{0.03\linewidth}\includegraphics[width=0.48\linewidth]{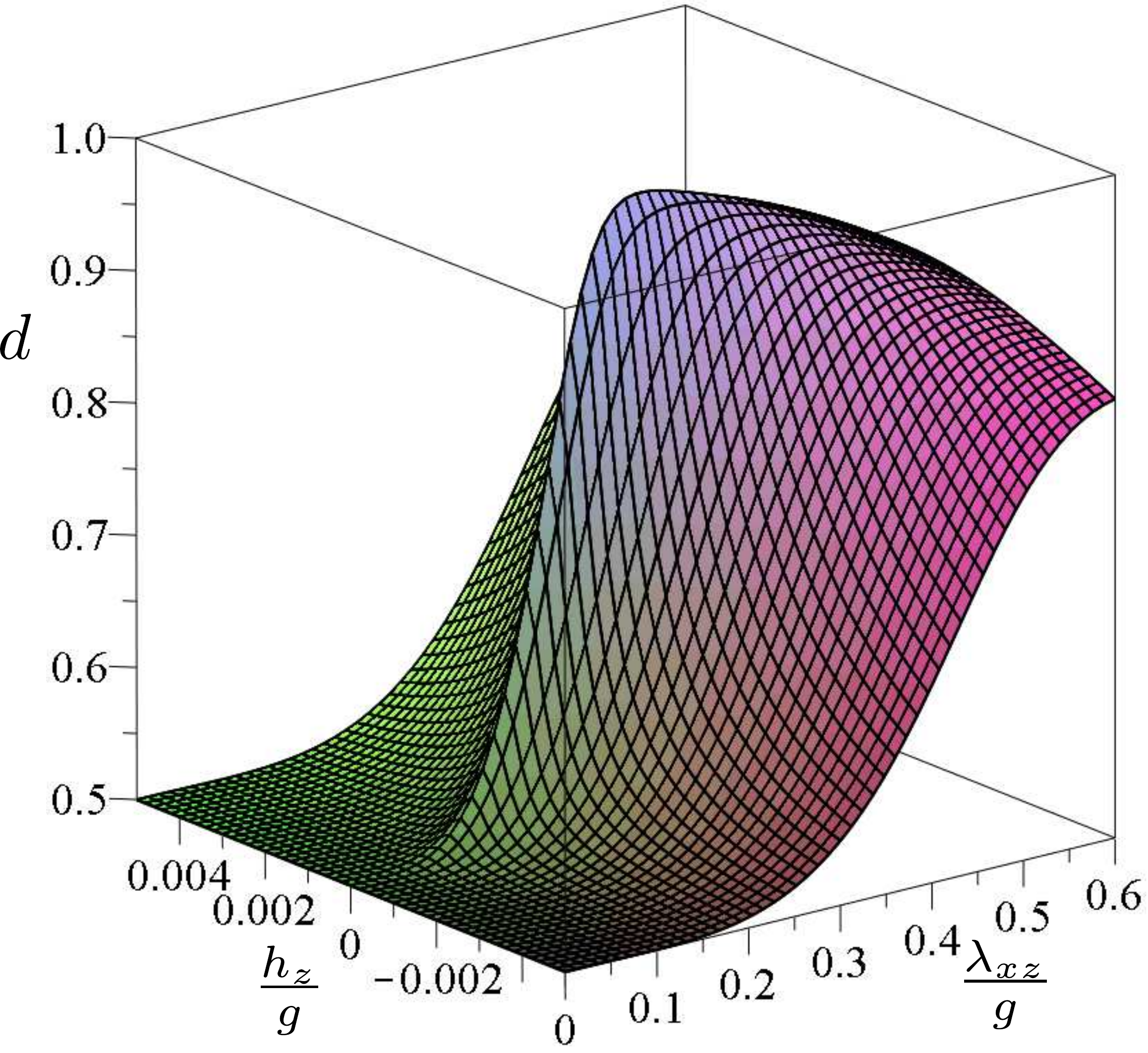}
\par\end{centering}

\caption{Energy gap $\Delta E_{z}=E_{\ket{1_{z}}}-E_{\ket{0_{z}}}$ (top left),
fidelity per site $d=|\braket{1_{z}|R_{z}|+}|^{2}$ of the first excited
state (top right), and fidelity per site $d$ (Eq.~\ref{eq:Fidelity with hz})
for the temperatures $T=0.0001\, g/k_{B}$ and $T=0.001\, g/k_{B}$
(bottom left and right) for the case $H_{\rm z}$ of a perturbation with an external
$Z$-field.\label{fig:Energy-gap-} }

\end{figure}

The energy gap $\Delta E_{z}=E_{\ket{1_{z}}}-E_{\ket{0_{z}}}$, the
fidelity $d=|\braket{1_{z}|R_{z}|+}|^{2}$ of the first excited state
, and the fidelity per site $d$ for the temperatures $T=0.0001\, g/k_{B}$
and $T=0.001g/k_{B}$ are shown in Fig.~\ref{fig:Energy-gap-}.
 By comparing with the fidelity per site for $T=0$ (shown in the
main paper) one sees, that for finite $h_{z}$ thermal fluctuations
play a minor role, as the gap is increased by the field.

\subsection*{Perturbation method by Takahashi}

To calculate the ground-state fidelity per site of the TFIM on a square
lattice with the polarized state, the fidelity per site of $H_{zz}$
with the logical cluster state, and the energy gap of $H_{zz}$ as
series expansions, we used a perturbation method introduced by Takahashi
\cite{0022-3719-10-8-031}. In terms of this method a Hamiltonian
$H=H_{0}+V$ is transferred by a base transformation $\Gamma$ into
an effective Hamiltonian $H_{{\rm eff}}$, so that it only acts in one of
the $H_{\ensuremath{0}}$ eigenspaces, while the spectral properties
are preserved.

Let $P$ be the projector into a $H_{0}$ eigenspace $L$ with eigenvalue
$E_{L}^{(0)}$ and $\bar{P}$ be the projector into the according
perturbed $H$ eigenspace $\bar{L}$. The Operator $\Gamma$ is then
defined by
\[
\Gamma:=\bar{P}P(P\bar{P}P)^{-\frac{1}{2}},
\]
with $\Gamma^{\dagger}\Gamma=P$. The projector $\bar{P}$ can be
expressed as
\begin{equation}
\bar{P}=P-\sum_{n=1}^{\infty}\sum_{k_{1}+k_{2}+...+k_{n+1}=n,\, k_{i}\geq0}S^{k_{1}}VS^{k_{2}}V...VS^{k_{n+1}}\label{eq:einsetzformel1}
\end{equation}
with $S^{0}:=-P$ and $S^{k}:=\left(\frac{1-P}{E_{L}^{(0)}-H_{0}}\right)^{k}$.
Together with
\begin{equation}
(P\bar{P}P)^{-\frac{1}{2}}=P+\sum_{n=1}^{\infty}\frac{(2n-1)!!}{(2n)!!}(P(P-\bar{P})P)^{n}\label{eq:einsetzformel2}
\end{equation}
it is now possible to express $\Gamma$ order by order as a power
series of $V$. Up to order 3 the coefficients of $\Gamma=\sum\Gamma^{(i)}$
are given by
\[
\Gamma^{(0)}=P,
\]

\[
\Gamma^{(1)}=SVP,
\]
\[
\Gamma^{(2)}=SVSVP-S^{2}VPVP-\frac{1}{2}PVS^{2}VP,
\]
\begin{eqnarray*}
\Gamma^{(3)} & = & S^{3}VPVPVP-S^{2}VSVPVP-SVS^{2}VPVP\\
 &  & +\frac{1}{2}PVS^{3}VPVP-S^{2}VPVSVP+SVSVSVP\\
 &  & -\frac{1}{2}PVS^{2}VSVP-\frac{1}{2}SVPVS^{2}VP\\
 &  & -\frac{1}{2}PVSVS^{2}VP+\frac{1}{2}PVPVS^{3}VP.
\end{eqnarray*}

Within the context of this work we calculated the coefficients up
to order 13 (7030571 coefficients).

We now transform the Schrödinger equation into the effective basis
\begin{eqnarray*}
H\ket{\psi_{i}} & = & E_{i}\ket{\psi_{i}}\\
\Rightarrow H_{{\rm eff}}\Gamma^{\dagger}\ket{\psi_{i}} & = & E_{i}\Gamma^{\dagger}\ket{\psi_{i}}
\end{eqnarray*}
with $H_{{\rm eff}}:=\Gamma^{\dagger}H\Gamma$. Every eigenvector
$\ket{\psi_{i}}$ of $H$ in $\bar{L}$ therefore corresponds to an
eigenvector of $H_{{\rm eff}}$ in $L$ with the same eigenvalue.
It is possible to calculate the perturbed eigenvector $\ket{\psi_{i}}$
with the base transformation $\Gamma$ from the according $H_{{\rm eff}}$
eigenvector $\Gamma^{\dagger}\ket{\psi_{i}}$.

To express the effective Hamiltonian $H_{{\rm eff}}=\sum H_{{\rm eff}}^{(i)}$
as a power series of $V$ we exploit that $[H,\bar{P}]=0$
\begin{eqnarray*}
H_{{\rm eff}} & = & (P\bar{P}P)^{-\frac{1}{2}}P\bar{P}H\bar{P}P(P\bar{P}P)^{-\frac{1}{2}}\\
 & = & (P\bar{P}P)^{-\frac{1}{2}}P(H_{0}+V)\bar{P}P(P\bar{P}P)^{-\frac{1}{2}}\\
 & = & E_{L}^{(0)}P+(P\bar{P}P)^{-\frac{1}{2}}PV\bar{P}P(P\bar{P}P)^{-\frac{1}{2}}.
\end{eqnarray*}
Together with Eqs.~\ref{eq:einsetzformel1}--\ref{eq:einsetzformel2}
one finds up to order 4
\[
H_{{\rm eff}}^{(0)}=E_{L}^{(0)}P,
\]
\[
H_{{\rm eff}}^{(1)}=PVP,
\]
\[
H_{{\rm eff}}^{(2)}=PVSVP,
\]
\[
H_{{\rm eff}}^{(3)}=PVSVSVP-\frac{1}{2}PVPVS^{2}VP-\frac{1}{2}PVS^{2}VPVP,
\]
\begin{eqnarray*}
H_{{\rm eff}}^{(4)} & = & PVSVSVSVP+\frac{1}{2}PVPVPVS^{3}VP-\frac{1}{2}PVPVSVS^{2}VP\\
 &  & -\frac{1}{2}PVPVS^{2}VSVP-\frac{1}{2}PVSVPVS^{2}VP-\frac{1}{2}PVSVS^{2}VPVP\\
 &  & -\frac{1}{2}PVS^{2}VPVSVP-\frac{1}{2}PVS^{2}VSVPVP+\frac{1}{2}PVS^{3}VPVPVP.
\end{eqnarray*}

Within the context of this work we calculated the coefficients up
to order 14 (5394321 coefficients) and for the special case $PVP=0$
up to order 17 (2490673 coefficients).

\subsection*{Application of Takahashi's method to the energy gap}

We applied this perturbation method to calculate the energy gap
between the ground state of $H_{zz}$ and its first excited state
up to order 5. We start from the representation of $H$ in the basis
$\Sigma^{{\rm diag}}$ and we define 
\[
H_{0}:=H^{{\rm diag}},
\]
\[
V:=-\lambda_{zz}\sum_{m\in\mathcal{M}}R_{\mathcal{L}}^{\dagger}Z_{m(1)}Z_{m(2)}R_{\mathcal{L}}.
\]

A method to calculate series expansions for properties of excited
states is based on a work by Gelfand \cite{MartinP199611}. The idea
is to interpret excitations on the lattice as quasi-particles and
to describe their dynamics. In this quasi-particle picture, the ground
state $\ket{\psi_{H}}$ of $H^{{\rm diag}}$ corresponds to the vacuum
state and the state $\ket{\psi_{{\rm EX}}^{(\mu)}}$ corresponds to
a state with a single particle at lattice site $\mu=(x,y)$ where $(x,y)$ are Cartesian coordinates. By switching
on the perturbation part, dynamics between quasi-particles are induced.
Using the effective form $H_{{\rm eff}}$ of $H_{zz}$ acting solely
in the space of the first excited states, we define the hopping elements
\[
t_{x,y}:=\braket{\psi_{{\rm EX}}^{{(a+x,b+y)}}|H_{{\rm eff}}|\psi_{{\rm EX}}^{{(a,b)}}}\quad a,b,x,y\in\mathbb{Z}.
\]
To this end, it is important that the perturbation $V$ is a sum of local operators each linking two neareast-neighbor lattice sites. A linked process is then defined as a linked operator sequence of these local operators. The linked-cluster theorem \cite{goldstone1957,springerlink:10.1007/BF01119617} states that only linked processes contribute to the one-particle amplitudes. 
It is therefore possible to retrieve the hopping elements in a given order (in
our case order 5) in the thermodynamic limit, although the calculations
are limited to finite systems.

To achieve this the finite systems
have to be constructed large enough to contain all linked processes
of up to 5 bonds. It also follows, that $|x|+|y|>5\Rightarrow t_{x,y}=0$,
as there is no order 5 process linking the lattice sites $(a,b)$
and $(a+x,b+y)$ for this case.

\begin{figure}[p]
\begin{centering}
\includegraphics[width=0.48\linewidth]{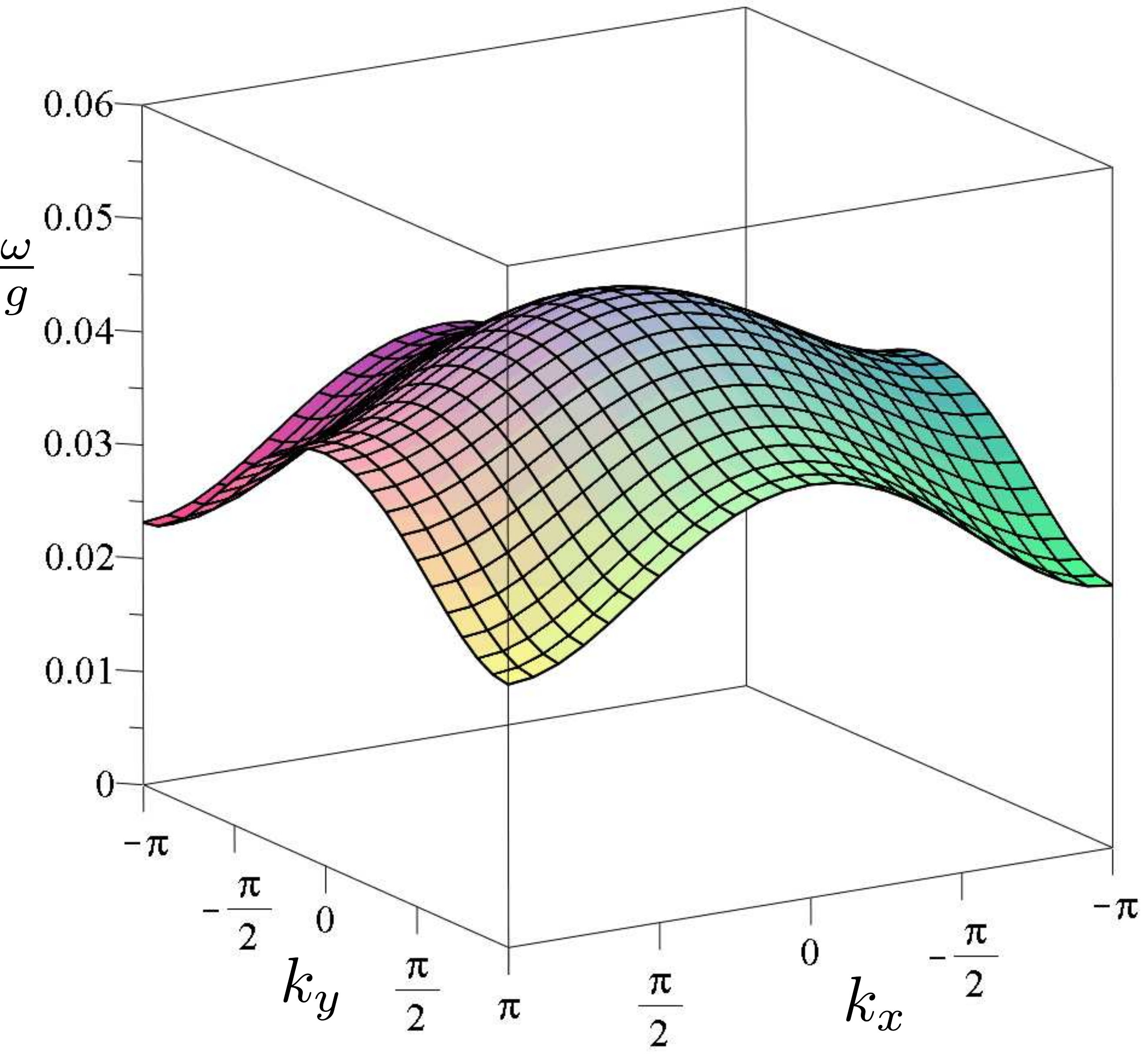}\hspace{0.03\linewidth}\includegraphics[width=0.48\linewidth]{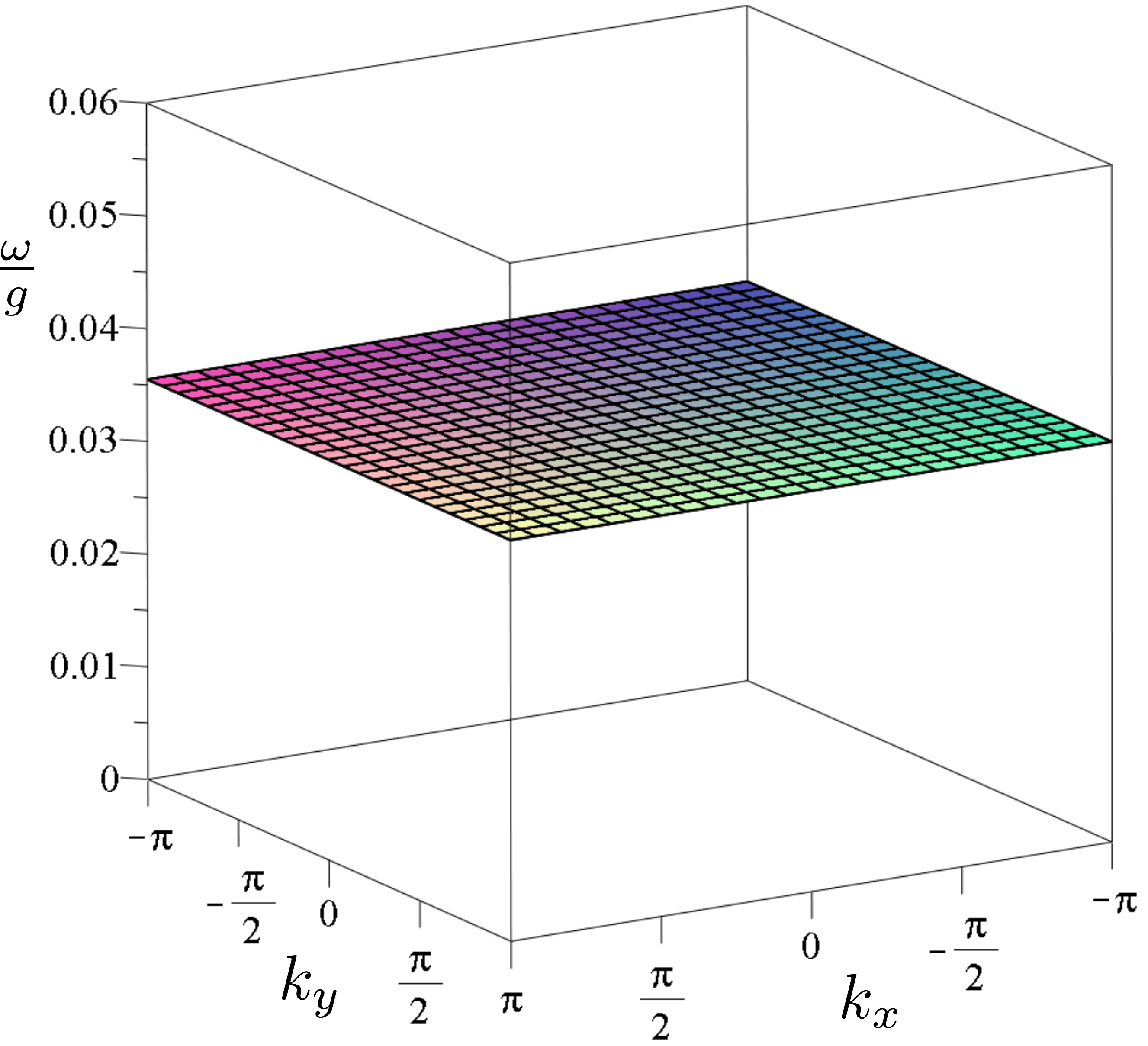}\bigskip{}
\includegraphics[width=0.48\linewidth]{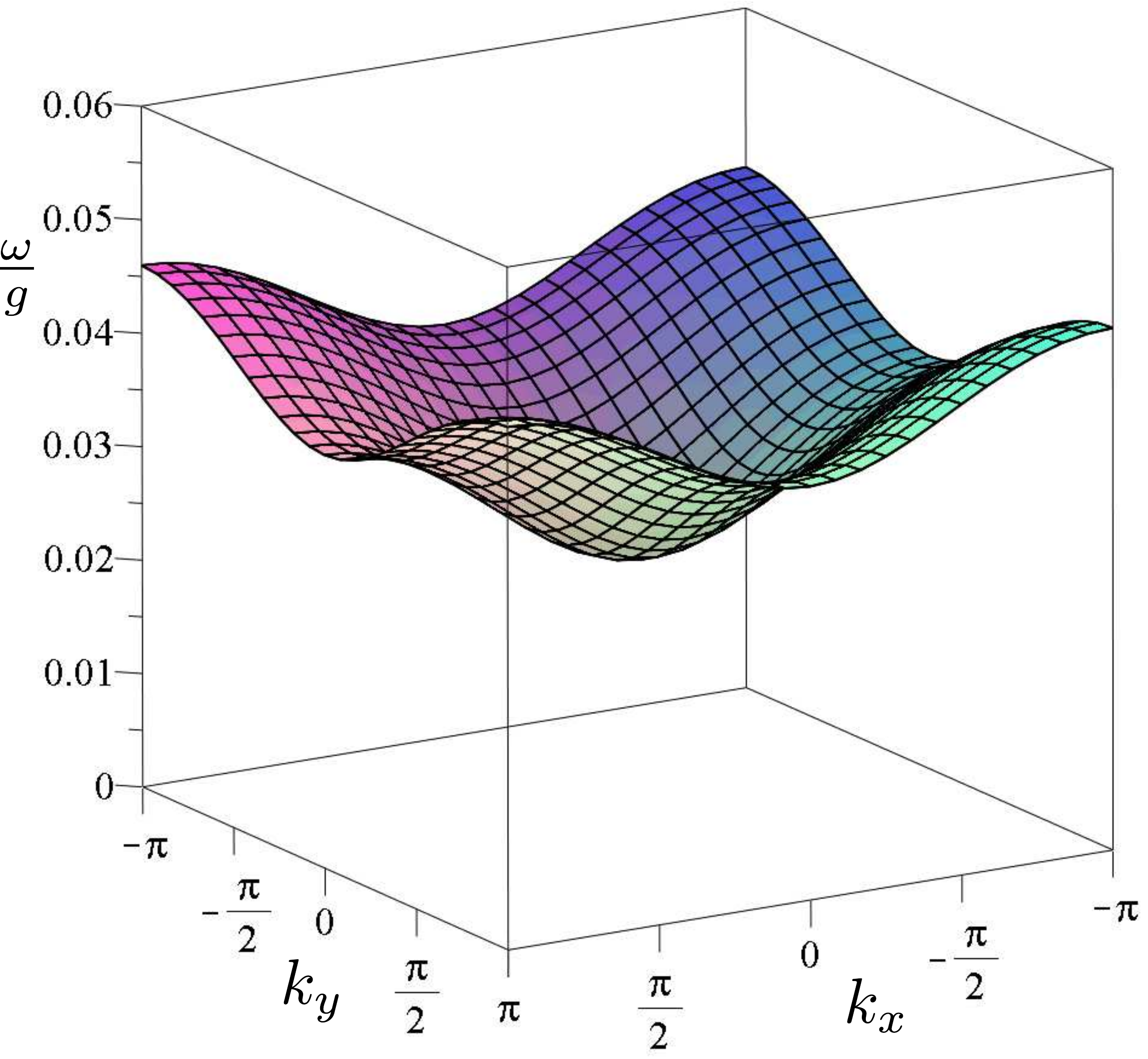}\hspace{0.03\linewidth}\includegraphics[width=0.48\linewidth]{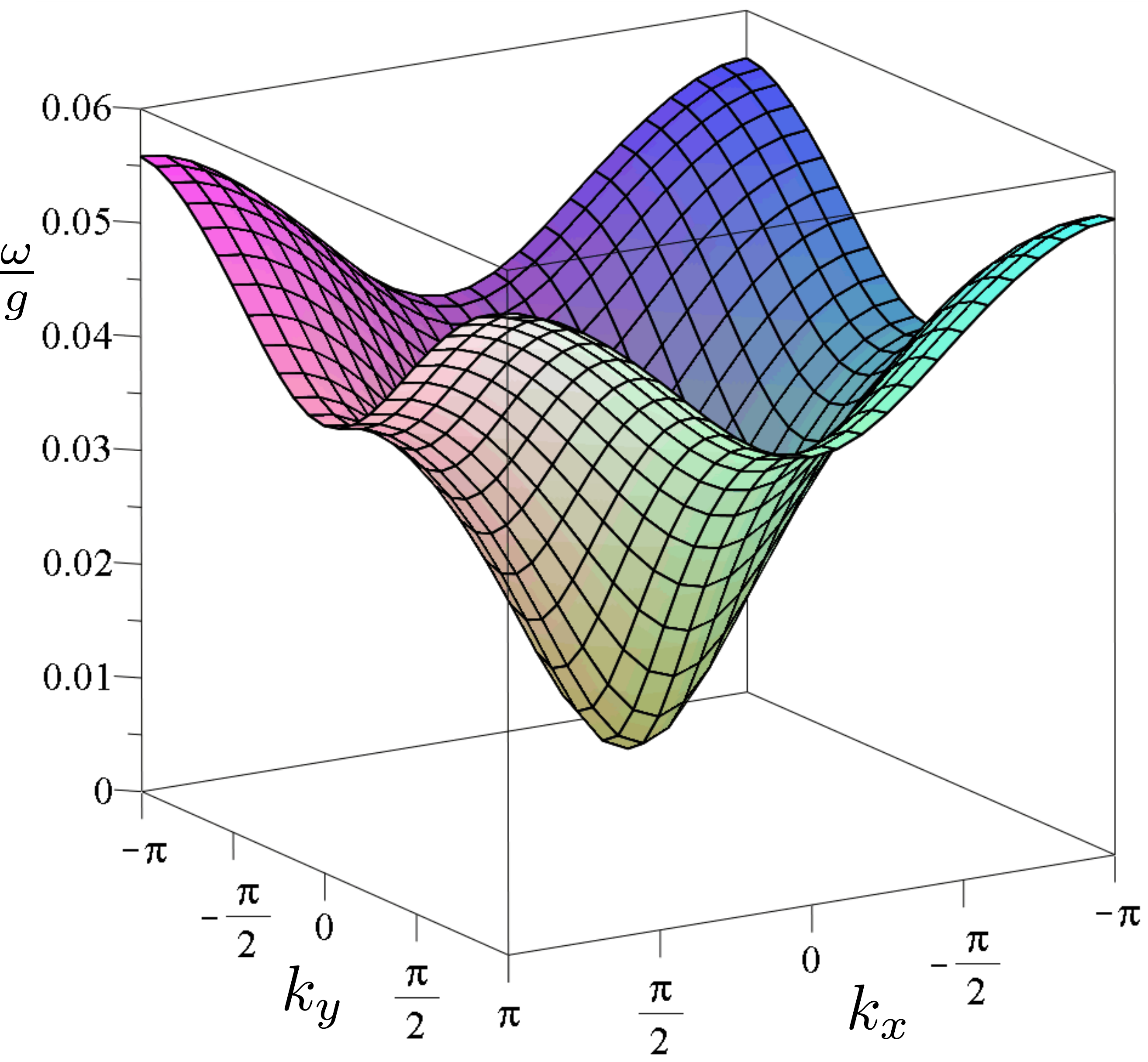}
\par\end{centering}

\caption{Dispersion of the first excited mode of $H_{zz}$ for $\lambda_{xz}=0.5\, g$
up to order 5 for different values of $\lambda_{zz}$: $\lambda_{zz}=-0.003\, g$ (top left), $\lambda_{zz}=0$ (top right), $\lambda_{zz}=0.003\, g$ (bottom left),
and $\lambda_{zz}=0.006\, g$ (bottom right).\label{fig:Dispersion-of-the}}

\end{figure}

\begin{figure}
\begin{centering}
\includegraphics[width=0.7\columnwidth]{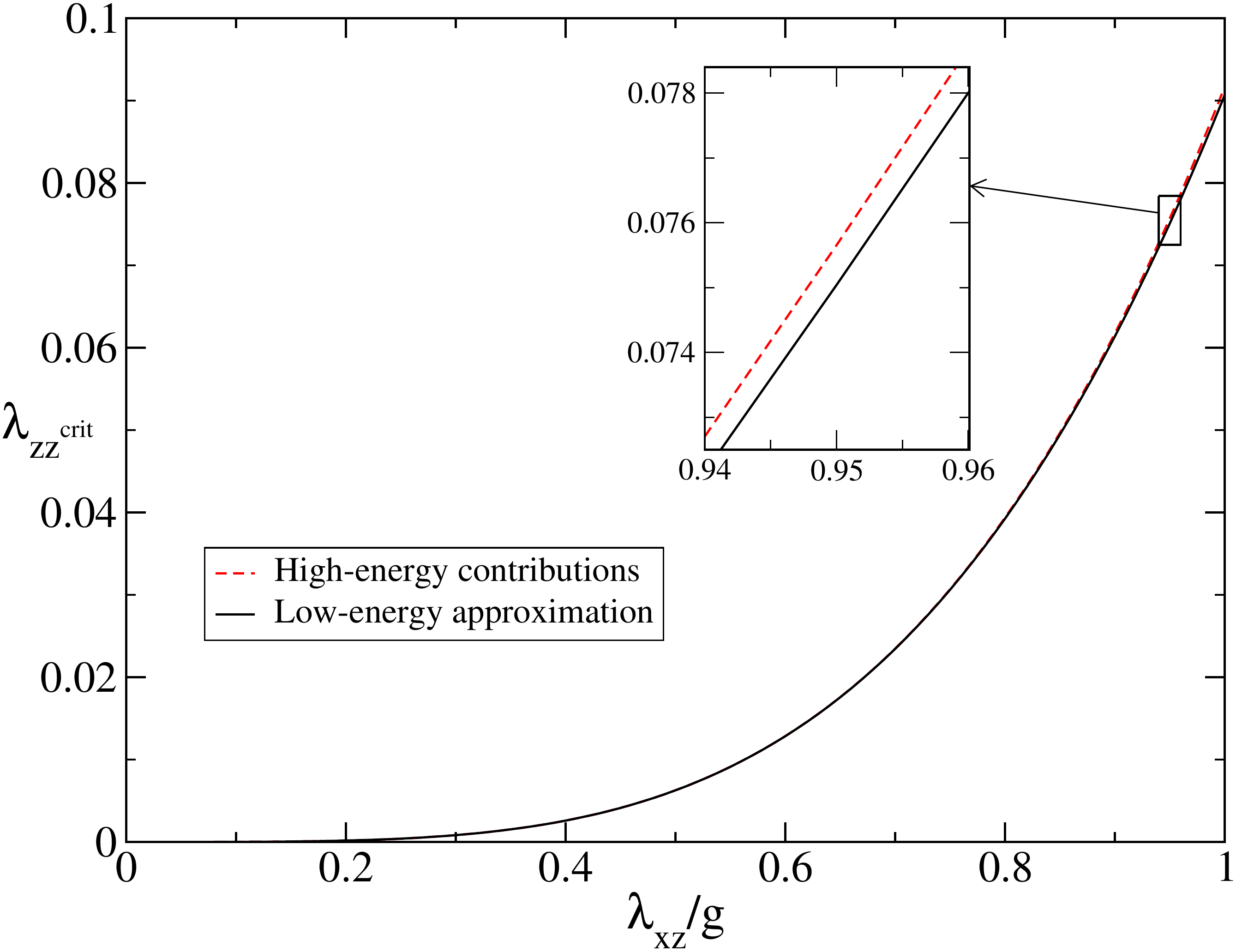}
\par\end{centering}

\caption{Critical $\lambda_{zz}|_{{\rm crit}}$ in dependence of $\lambda_{xz}$
calculated with the low-energy approximation (black solid line) and corrected by high-energy
contributions (red dashed line). The high-energy corrections are calculated by a dlogPadé~{[}2,2{]}
approximation of the order 5 series of the energy gap.\label{fig:Critical--in}}

\end{figure}

The calculation can further be simplified:
\begin{enumerate}
\item For symmetry reasons it is
\[
t_{x,y}=t_{-x,y}=t_{x,-y}=t_{-x,-y}\quad,
\]
so only one calculation is necessary for such hopping elements.
\item An operator $Z_{m(1)}Z_{m(2)}$ of $V$ anti-commutes with $K_{m(1)}$
and $K_{m(2)}$. It follows that the application of such an operator inverts
both $K_{\mu}$ eigenvalues of the two neighboring lattice sites. To contribute
to a hopping element, each lattice site must therefore be touched an even number
of times by such operators. The only exception to this rule are the lattice sites $(a,b)$ and $(a+x,b+y)$ for $(x,y)\neq(0,0)$.
They must be touched by an uneven number of operators. This allows to
further decrease the size of the finite systems. Examples for some
hopping elements can be found in Fig.~\ref{fig:Examples-for-finite}.
\begin{figure}
\begin{centering}
\includegraphics[width=0.8\columnwidth]{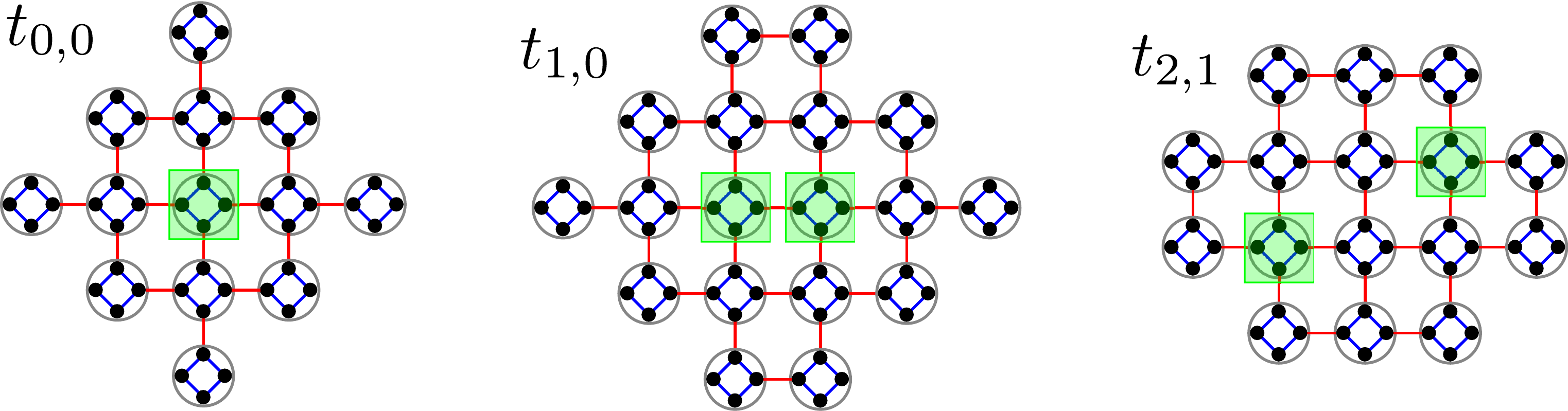}
\par\end{centering}

\caption{Examples for finite systems to calculate one-particle hopping elements $t_{x,y}$ from the lattice site $(a,b)$ to $(a+x,b+y)$ or vice versa are shown. These two lattice sites are highlighted by a green (gray) background. The finite systems are constructed such that all connected graphs of up to five bonds (a single bond can be counted multiple times) are contained, where each lattice site is touched by an even number of bonds. The only exception to this rule are the lattice sites $(a,b)$ and $(a+x,b+y)$ which must be touched by an odd number of bonds for $(x,y)\neq(0,0)$. The hopping elements of this systems equal the hopping elements of an infinite system up to order 5. \label{fig:Examples-for-finite}}
\end{figure}

\item If $h$ is an operator of the power series representation of $H_{{\rm eff}}$
one can calculate its contribution to a hopping element by calculating
$h\ket{\psi_{{\rm EX}}^{{(a,b)}}}$ and multiplying the result with
$\bra{\psi_{{\rm EX}}^{{(a+x,b+y)}}}$. But to save time and memory
it is useful to split the coefficient $h=h_{l}h_{r}$ into two parts
with roughly the same number of $V$ Operators. The contribution is
then calculated by multiplying the two vectors $h_{r}\ket{\psi_{{\rm EX}}^{{(a,b)}}}$
and $\bra{\psi_{{\rm EX}}^{{(a+x,b+y)}}}h_{l}$.
\end{enumerate}
From the hopping elements we calculate the dispersion of the first
excited mode with a Fourier transformation. For the square lattice
one finds
\begin{equation}
\omega(k_{x},k_{y})=t_{0,0}-E_{0}+\sum_{(x,y)\neq(0,0)}t_{x,y}\, cos(k_{x}x+k_{y}y),\label{eq:dispersion}
\end{equation}
where $E_{0}$ is the ground-state energy of the finite lattice used
to calculate $t_{0,0}$. The dispersions for the case $\lambda_{xz}=0.5\, g$
and for different $\lambda_{zz}$ are shown in Fig.~\ref{fig:Dispersion-of-the}.
 The minimal excitation energy (the gap) is at $k_{x}=k_{y}=\pi$ for negative
$\lambda_{zz}$ and at $k_{x}=k_{y}=0$ for positive $\lambda_{zz}$.
 The coefficients
of the series expansion of the gap up to order 5 are shown in Tab.~\ref{tab:Koeffizienten-der-Reihenentwickl gap}
for different $\lambda_{xz}$. The first-order coefficients equal
the coefficient for the TFIM gap. The absolute value of all other
coefficients are decreased for finite $\lambda_{xz}$. The actual
value of the critical $\lambda_{zz}|_{{\rm crit}}$ is therefore slightly
larger then the critical point calculated using the low-energy approximation
(see Fig.~\ref{fig:Critical--in}).

\begin{table}
\begin{centering}
\begin{tabular}{|c|c|c|c|c|c|}
\hline 
 & $c_{1}\, f_{1}$ & $c_{2}\, f_{2}$ & $c_{3}\, f_{3}$ & $c_{4}\, f_{4}$ & $c_{5}\, f_{5}$\tabularnewline
\hline 
\hline 
TFIM ($\lambda_{xz}\rightarrow0$) & $-4$ & $-2$ & $-3$ & $-4.5$ & $-11$\tabularnewline
\hline 
$\lambda_{xz}=0.1\, g$ & $-4.000$ & $-2.000$ & $-3.000$ & $-4.500$ & $-11.00$\tabularnewline
\hline 
$\lambda_{xz}=0.2\, g$ & $-4.000$ & $-2.000$ & $-3.000$ & $-4.500$ & $-11.00$\tabularnewline
\hline 
$\lambda_{xz}=0.3\, g$ & $-4.000$ & $-2.000$ & $-3.000$ & $-4.500$ & $-11.00$\tabularnewline
\hline 
$\lambda_{xz}=0.4\, g$ & $-4.000$ & $-2.000$ & $-3.000$ & $-4.500$ & $-11.00$\tabularnewline
\hline 
$\lambda_{xz}=0.5\, g$ & $-4.000$ & $-1.998$ & $-2.998$ & $-4.498$ & $-10.99$\tabularnewline
\hline 
$\lambda_{xz}=0.6\, g$ & $-4.000$ & $-1.992$ & $-2.994$ & $-4.493$ & $-10.97$\tabularnewline
\hline 
$\lambda_{xz}=0.7\, g$ & $-4.000$ & $-1.977$ & $-2.983$ & $-4.480$ & $-10.93$\tabularnewline
\hline 
$\lambda_{xz}=0.8\, g$ & $-4.000$ & $-1.945$ & $-2.959$ & $-4.453$ & $-10.83$\tabularnewline
\hline 
$\lambda_{xz}=0.9\, g$ & $-4.000$ & $-1.888$ & $-2.913$ & $-4.402$ & $-10.64$\tabularnewline
\hline 
$\lambda_{xz}=1.0\, g$ & $-4.000$ & $-1.798$ & $-2.838$ & $-4.319$ & $-10.36$\tabularnewline
\hline 
\end{tabular}
\par\end{centering}

\caption{Coefficients of the energy gap $\Delta E+c_{1}\frac{\lambda_{zz}}{g}+c_{2}(\frac{\lambda_{zz}}{g})^{2}+c_{3}(\frac{\lambda_{zz}}{g})^{3}+c_{4}(\frac{\lambda_{zz}}{g})^{4}+c_{5}(\frac{\lambda_{zz}}{g})^{5}$
of $H_{zz}$ up to order 5. The first row shows the coefficients for
the TFIM (limit $\lambda_{xz}\rightarrow0$). It is $f_{i}:=\frac{2}{\Delta E}\,(\frac{\Delta E}{2\, g\, c^{2}})^{i}$.\label{tab:Koeffizienten-der-Reihenentwickl gap}}

\end{table}

\subsection*{Application of Takahashi's method to the ground-state fidelity}

We have applied the perturbation method to calculate the ground-state
fidelity per site of $H_{zz}$ with the logical cluster state up to
order 4 and the ground-state fidelity per site of a TFIM on a square
lattice with the polarized state up to order 12. For $H_{zz}$,
we again use the basis $\Sigma^{{\rm diag}}$ and define
\[
H_{0}:=H^{{\rm diag}},
\]
\[
V:=-\lambda_{zz}\sum_{m\in\mathcal{M}}R_{\mathcal{L}}^{\dagger}Z_{m(1)}Z_{m(2)}R_{\mathcal{L}}.
\]
For the TFIM, we use
\[
H_{0}:=-\frac{\Delta E}{2}\sum_{\mu\in\mathcal{L}}\tilde{Z}_{\mu},
\]
\[
V:=-\lambda_{zz}c^{2}\sum_{m\in M}\tilde{X}_{m(1)}\tilde{X}_{m(2)}.
\]
The ground state $\ket{\psi_{H}}=\bigotimes_{\mu\in\mathcal{L}}\ket{0}_{\mu}$
of $H_{0}$ is in both cases unique and therefore also the only eigenvector
of the according $H_{{\rm eff}}$. The wave functions of the perturbed
ground states can therefore be calculated by applying the base transformation
$\Gamma$ to $\bigotimes_{\mu\in\mathcal{L}}\ket{0}_{\mu}$. For the
ground-state fidelity per site one finds 
\[
d_{H_{zz}}(\ket{\psi_{{\rm CS}}},\ket{\psi_{H_{zz}}})=\lim_{N\rightarrow\infty}\sqrt[N]{|\braket{\psi_{{\rm CS}}^{{\rm diag}}|\Gamma|0}_{\mathcal{L}}|^{2}},
\]
\[
d_{{\rm TFIM}}(\ket{0}_{\mathcal{L}},\ket{\psi_{{\rm TFIM}}})=\lim_{N\rightarrow\infty}\sqrt[N]{|\braket{0|_{\mathcal{L}}\Gamma|0}_{\mathcal{L}}|^{2}}.
\]
Again it is possible to exploit the linked-cluster theorem to limit
the calculation to finite systems. For example it is sufficient to
calculate on a $5\times5$ lattice with periodic boundary conditions
(system lies on a torus) for the order 4 calculation, or a $13\times13$
lattice for the order 12 calculation respectively.

\begin{figure}
\begin{centering}
\includegraphics[width=0.7\columnwidth]{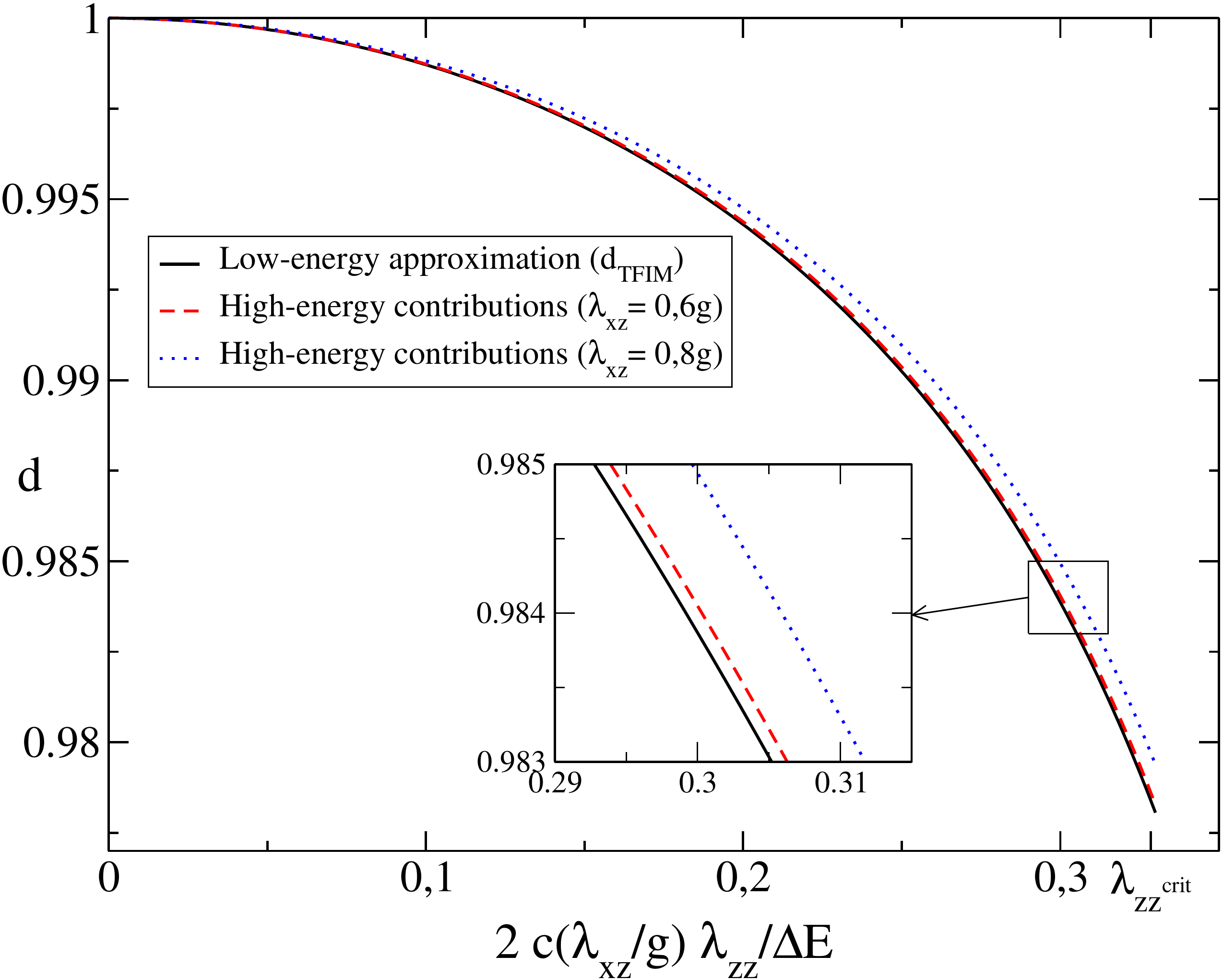}
\par\end{centering}

\caption{Ground-state fidelity per site $d$ of $H_{zz}$ calculated with the low-energy
approximation ($d_{{\rm TFIM}}$, shown as solid line) and corrected by high-energy contributions
for different $\lambda_{xz}$ (dashed and dotted lines). The low-energy fidelity is calculated
up to order 12 and the corrections are calculated up to order 4.\label{fig:Ground-state-fidelity}}

\end{figure}

The calculation can further be simplified.
\begin{enumerate}
\item Analog to the second point of the last section only processes can
contribute, where each lattice site is touched by an even number of operators located on the bonds. It is easy to see, that this is only possible if the overall
number of bonds is even. For the calculations only coefficients of
even order must be considered.
\item The $5\times5$ lattice necessary for the order 4 calculation can
be reduced to a $3\times5$ lattice by bounding it periodically into
a ``brick wall'' structure (see Fig.~\ref{fig:Periodic-lattice-with}).
\begin{figure}
\begin{centering}
\includegraphics[width=0.7\columnwidth]{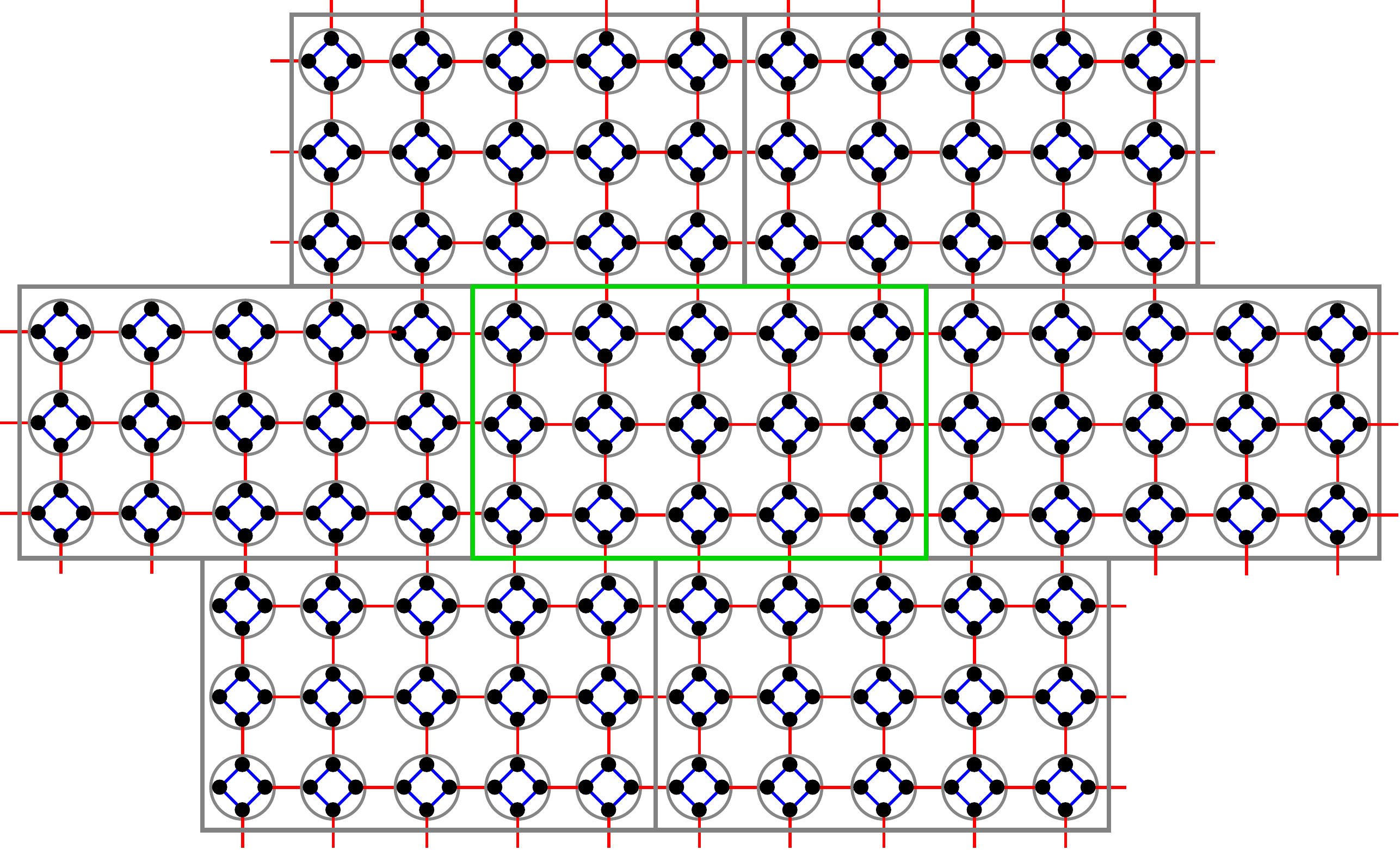}
\par\end{centering}

\caption{Periodic lattice with ``brick wall'' boundary conditions. All linked
processes of at most 4 bonds cannot form closed loops crossing two
opposed lattice boundaries. The order 4 fidelity per site of this
lattice is therefore identically to the fidelity in the thermodynamic
limit.\label{fig:Periodic-lattice-with}}
\end{figure}

\item To reduce the size of the $13\times13$ lattice for the order 12 calculation
another technique was used. The coupling parameter $\lambda_{zz}$
was replaced by the parameters $\lambda_{|}$ and $\lambda_{-}$,
so that vertical bonds are associated with $\lambda_{|}$ and horizontal
bonds are associated with $\lambda_{-}$. Using the rotational symmetry
of the lattice the calculation of the order 12 coefficient $c_{12}$
is reduced to the calculation of the $\lambda_{|}$ and $\lambda_{-}$
coefficients
\[
c_{12}\lambda_{zz}^{12}=c_{12}^{(0,12)}(\lambda_{|}^{12}+\lambda_{-}^{12})+c_{12}^{(2,10)}(\lambda_{|}^{2}\lambda_{-}^{10}+\lambda_{|}^{10}\lambda_{-}^{2})+c_{12}^{(4,8)}(\lambda_{|}^{4}\lambda_{-}^{8}+\lambda_{|}^{8}\lambda_{-}^{4})+c_{12}^{(6,6)}\lambda_{|}^{6}\lambda_{-}^{6}.
\]
For the separate calculations of the coefficients $c_{12}^{(0,12)}$,
$c_{12}^{(2,10)}$, $c_{12}^{(4,8)}$ and $c_{12}^{(6,6)}$ periodically
bounded systems of size $1\times13$, $3\times11$, $5\times9$ and
$7\times7$ are sufficient.
\item It is $\bra{0}_{\mathcal{L}}S=0$. For the calculation of $\braket{0|_{\mathcal{L}}\Gamma|0}_{\mathcal{L}}$
all coefficients with an $S$-operator at the left side can therefore
be left out.
\item Analog to point 3 of the last section it is possible to reduce the
calculation of the contribution $\braket{0|_{\mathcal{L}}\gamma|0}_{\mathcal{L}}$
of a coefficient $\gamma=\gamma_{l}\gamma_{r}$ of the power series
of $\Gamma$ to the calculation of $\bra{0}_{\mathcal{L}}\gamma_{l}$
and $\gamma_{r}\ket{0}_{\mathcal{L}}$ and their product. In principle
this can be also applied to the calculation of $\braket{\psi_{{\rm CS}}^{{\rm diag}}|\gamma|0}_{\mathcal{L}}$.
Due to the size of the representation of $\bra{\psi_{{\rm CS}}^{{\rm diag}}}$
in the basis $\Sigma^{{\rm diag}}$ this does not save time or memory.
It is better to calculate $\gamma\ket{0}_{\mathcal{L}}$ and multiply
it with $\bra{\psi_{{\rm CS}}^{{\rm diag}}}$ directly.
\item For symmetry reasons the contribution of a process only depends on
the shape of the process and not on its position on the lattice. It
is therefore possible to limit one of the $V$-operators (advisably
the first) to a single and fixed bond. The resulting value is then
the fidelity of the system divided by the number of bonds.
\end{enumerate}
The resulting ground-state fidelity per site of the TFIM on a square
lattice with the polarized state is given by 
\begin{eqnarray*}
d_{{\rm TFIM}} & = & 1-\frac{1}{8}\lambda^{2}-{\frac{93}{256}}\lambda^{4}-{\frac{2961}{2048}}{\it \lambda}^{6}-{\frac{243005}{32768}}{\it \lambda}^{8}\\
 &  & -{\frac{812949139}{18874368}}{\it \lambda}^{10}-{\frac{17716040461601}{65229815808}}{\it \lambda}^{12}.
\end{eqnarray*}
The coefficients of the ground-state fidelity per site of $H_{zz}$
with the logical cluster state are shown in Tab.~\ref{tab:Koeffizienten-der-Reihenentwickl fidelity}.
One sees, that the absolute values of the coefficients are decreased
for finite $\lambda_{xz}$. The actual ground-state fidelity per site
of $H_{zz}$ is therefore slightly larger then the fidelity calculated
with the low-energy approximation. The fidelity $d_{{\rm TFIM}}$
and the fidelity of $H_{zz}$ for different $\lambda_{xz}$ are plotted
in Fig.~\ref{fig:Ground-state-fidelity}.

\begin{table}
\begin{centering}
\begin{tabular}{|c|c|c|}
\hline 
 & $c_{2}\cdot\frac{1}{|\braket{0|R|+}|^{2}}\cdot\left(\frac{\Delta E}{2\cdot g\cdot c^{2}}\right)^{2}$ & $c_{4}\cdot\frac{1}{|\braket{0|R|+_{L}}|^{2}}\cdot\left(\frac{\Delta E}{2\cdot g\cdot c^{2}}\right)^{4}$\tabularnewline
\hline 
\hline 
TFIM ($\lambda_{xz}\rightarrow0$) & $-0.125$ & $-0.3633$\tabularnewline
\hline 
$\lambda_{xz}=0.1\, g$ & $-0.1250$ & $-0.3633$\tabularnewline
\hline 
$\lambda_{xz}=0.2\, g$ & $-0.1250$ & $-0.3633$\tabularnewline
\hline 
$\lambda_{xz}=0.3\, g$ & $-0.1250$ & $-0.3633$\tabularnewline
\hline 
$\lambda_{xz}=0.4\, g$ & $-0.1249$ & $-0.3631$\tabularnewline
\hline 
$\lambda_{xz}=0.5\, g$ & $-0.1244$ & $-0.3627$\tabularnewline
\hline 
$\lambda_{xz}=0.6\, g$ & $-0.1232$ & $-0.3612$\tabularnewline
\hline 
$\lambda_{xz}=0.7\, g$ & $-0.1202$ & $-0.3572$\tabularnewline
\hline 
$\lambda_{xz}=0.8\, g$ & $-0.1144$ & $-0.3485$\tabularnewline
\hline 
$\lambda_{xz}=0.9\, g$ & $-0.1051$ & $-0.3321$\tabularnewline
\hline 
$\lambda_{xz}=1.0\, g$ & $-0.09228$ & $-0.3049$\tabularnewline
\hline 
\end{tabular}
\par\end{centering}

\caption{Coefficients of the ground-state fidelity per site $d_{H_{zz}}=|\braket{0|R|+}|^{2}+c_{2}(\frac{\lambda_{zz}}{g})^{2}+c_{4}(\frac{\lambda_{zz}}{g})^{4}$
of $H_{zz}$ with the logical cluster state for different $\lambda_{xz}$.
The first row shows the coefficients for the TFIM (limit $\lambda_{xz}\rightarrow0$).\label{tab:Koeffizienten-der-Reihenentwickl fidelity}}

\end{table}

\end{document}